\documentclass[12pt]{article}

\usepackage{newtxtext,newtxmath}

\usepackage{graphicx}

\usepackage[letterpaper,margin=1in]{geometry}

\renewenvironment{abstract}
	{\quotation}
	{\endquotation}

\date{}

\usepackage{url}
\usepackage[hidelinks]{hyperref}
\usepackage{caption}
\usepackage{float}
\usepackage{threeparttable}
\usepackage{xurl}
\usepackage{booktabs}

\newcommand{\vp}{\mathbf{p}}
\newcommand{\vw}{\mathbf{w}}
\newcommand{\vd}{\mathbf{d}}
\newcommand{\Ne}{N_{\mathrm{e}}}
\newcommand{\argmax}{\mathop{\mathrm{arg\,max}}}

\newcommand{\norm}[1]{\left\lVert#1\right\rVert}
\newcommand{\thickhline}{\noalign{\hrule height1pt}}

\def\scititle{
Stable and Accurate Orbital-Free DFT\\ Powered by Machine Learning 
}

\title{\bfseries \boldmath \textsf \scititle}

\author{
    \textsf{R.~Remme, T.~Kaczun, T.~Ebert, C.~A.~Gehrig, D.~Geng, G.~Gerhartz,} \and \textsf{M.~K.~Ickler, M.~V.~Klockow, P.~Lippmann, J.~S.~Schmidt, S.~Wagner,} \and
    \textsf{A.~Dreuw, F.~A.~Hamprecht$^\ast$} \and
    \textsf{\small Interdisciplinary Center for Scientific Computing (IWR), Heidelberg University,} \and
    \textsf{\small 69120 Heidelberg, Germany.} \\
	\textsf{\small$^\ast$Corresponding author. Email: fred.hamprecht@iwr.uni-heidelberg.de} \and
}

\begin{document} 

\maketitle

\begin{abstract} \bfseries \boldmath
    \fontsize{11}{9}

    \textsf{
    Hohenberg and Kohn have proven that the electronic energy and the one-particle
electron density can, in principle, be obtained by minimizing an energy functional with
respect to the density. While decades of theoretical work have produced increasingly
faithful approximations to this elusive exact energy functional, their accuracy is still
insufficient for many applications, making it reasonable to try and learn it empirically.
Using rotationally equivariant atomistic machine learning, we obtain for the first time a
density functional that, when applied to the organic molecules in QM9, yields energies
with chemical accuracy relative to the Kohn-Sham reference while also converging to
meaningful electron densities. Augmenting the training data with densities obtained
from perturbed potentials proved key to these advances. This work demonstrates that
machine learning can play a crucial role in narrowing the gap between theory and the
practical realization of Hohenberg and Kohn’s vision, paving the way for more efficient
calculations in large molecular systems.
    }
\end{abstract}

\newpage
\noindent
In a disarmingly simple proof, Hohenberg and Kohn showed~\cite{hohenberg1964inhomogeneous} that the electron density alone is sufficient to determine the ground state energy of a molecular system. The proof marked a radical departure from most prior work, which had recurred to the Schrödinger equation acting on a multielectron wave function to describe a system of interacting electrons. Of note, for $N_e$ electrons, a general wave function lives (omitting spin for simplicity) in $\mathbb{R}^{3N_e}$. Mean-field approximations such as Hartree-Fock reduce this to $N_e$ coupled one-electron wave functions, called orbitals, each living in $\mathbb{R}^3$. The electron density, on the other hand, is a \emph{single} function in $\mathbb{R}^3$, teasing the possibility of a profound simplification of the description of multielectron quantum systems. This promise is reinforced by the second Hohenberg-Kohn theorem enunciating the existence of a variational principle: Not only is there an energy functional that assigns an energy to each density; but the ground state electron density $\rho(\mathbf{r})$ is a minimizer of that functional. 

Sadly, the proof of existence is non-constructive. That is, we know that a universal energy functional $F[\rho]$ exists; but its exact form remains unknown except for simple special cases~\cite{weizsacker1935theorie, thomas1927calculation, fermi1928statistische} that do not cover most chemistries of general interest. 

This limitation led Kohn and Sham to re-introduce auxiliary one-electron wave functions for the sole reason that this representation would allow invoking the well-known quantum mechanical operator for the kinetic energy~\cite{kohn1965self}. The enormous practical success of the resulting Kohn-Sham density functional theory, or KS-DFT for short, makes it the most widely used quantum chemical method today and arguably is what turned theoretical chemistry into a practically applied discipline. Its success also crowded out methods relying on the electron density alone, now called orbital-free density functional theory (OF-DFT). 

Optimism that the latter can be made practical is founded on the ``nearsightedness of electronic matter'' which Kohn attributes to wave-mechanical destructive interference~\cite{kohn1996density}. Admittedly, we are also emboldened by what may be called the ``miracle of chemistry'': The empirical fact that chemists are able to make semi-quantitative predictions of stability and reactivity in their heads, even though the underlying systems are profoundly quantum mechanical. In other words, it is fair to assume some measure of locality and of well-behavedness of the unknown energy functional, at least for systems with a band gap. %

The lure of a simple but complete description of molecular ground states in terms of their electron density alone
has motivated an intense search for approximations with relatively few parameters \cite{yang1986molecules,wang1992nonlocal,lembarki1994fromxc,thakkar1992comparison,wang1999orbital,constantin2011semiclassical,karasiev2013nonempirical,luo2018simple,Witt_delRio_Dieterich_Carter_2018,shao2021revised,mi2023orbital}. While remarkably successful considering the complex nature of the kinetic energy, these are not yet sufficiently accurate for the quantitative prediction of outcomes in typical laboratory chemistry.

This motivates the search for machine learned functionals whose  significantly larger number of parameters promises greater expressivity
\cite{snyder2012finding,kulik2022roadmap,bogojeski2020quantum}: While promising, initial forays~\cite{yao2016kinetic,seino2018semi-local,fujinami2020orbital-free,ryczko2022toward} were still limited regarding accuracy and/or transferability of the learned functionals.
The equivariant KineticNet architecture~\cite{remme2023kineticnet} was the first to demonstrate chemical accuracy of a single learned functional across input densities and geometries of tiny molecules. 
This work was superseded by the milestone \textsc{M-OFDFT} pipeline~\cite{zhang2024overcoming}. Its representation of the density in terms of a linear combination of atomic basis functions~\cite{grisafi2018transferable, vergara2023efficient} is much more compact, compared to a grid. 
It %
first predicted molecular energies across the QM9 dataset~\cite{ruddigkeit2012enumeration, ramakrishnan2014quantum} of diverse organic molecules with up to nine second-period atoms with chemical accuracy with respect to the PBE/6-31G(2df,p) Kohn-Sham reference, a remarkable feat. Perhaps the biggest success of that work was its ability to extrapolate to larger molecules than trained on. 
One drawback was the model being fully non-local, i.e., each atom needs to exchange information with all others. While not a problem for molecules up to a few hundred atoms, this eventually becomes prohibitive for larger systems. %
Its foremost limitation, however, was that the learned functional did not afford variational density optimization: Following the energy gradient led to unphysical densities and energies, creating the need to engineer a method that picks an intermediate solution as the final prediction post hoc. 

\begin{figure}[t]
    \centering
    \includegraphics[width=\linewidth]{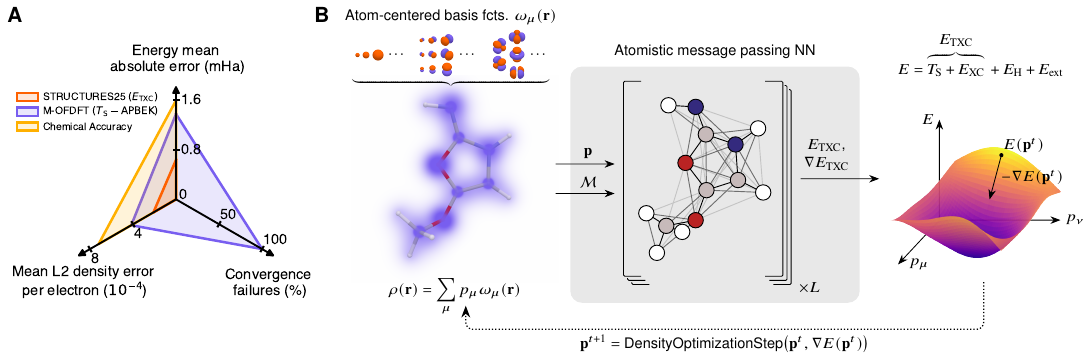}
    \caption{
    \textbf{The STRUCTURES25 pipeline enables converging orbital-free electron density optimization.}
    (\textbf{A}) Radar plot of total energy error, $L_2$ density error and percentage of convergence failures 
    evaluated on QM9. STRUCTURES25 was trained on our perturbed QM9 dataset. Both energy and density error are w.r.t.~PBE/6-31G(2df,p) ground state labels in the orbital-free density basis (Eq.~\ref{eq:density_basis}). 
    (\textbf{B}) The STRUCTURES25 pipeline takes a molecular graph $\mathcal{M}$ and a density represented by coefficients $\vp$ of atom-centered basis functions $\{\omega_\mu\}$ as input and predicts the target energy, $E_{\mathrm{TXC}}$. The gradient of the energy is obtained by automatic differentiation and used to iteratively find the ground state in density optimization.}
    \label{fig:overview}
\end{figure}

Learning a density functional which makes truly variational optimization possible has been our main objective, and below we describe how this objective is achieved (Fig.~\ref{fig:overview}A) by letting the model learn from physically plausible, but %
more varied, and more evenly distributed training data. %
A secondary objective has been to address the scaling of computational cost with size. We demonstrate good extrapolation to larger systems with a guaranteed field of view that does not need to grow with the system of interest. %

\section*{Variational density optimization}

As illustrated in Fig.~\ref{fig:overview}B, a suitable machine learning architecture can be used to map an electron density (here represented as a coefficient vector $\mathbf{p}$ of atom-centered basis functions) for a given molecular constitution and geometry $\mathcal{M}$ to an energy estimate $E(\mathbf{p}, \mathcal{M})$. A~``variational'' density optimization then takes gradient descent steps on this energy surface to iteratively update the density coefficients.

\begin{figure}[t!]
    \centering
    \includegraphics[width=\linewidth]{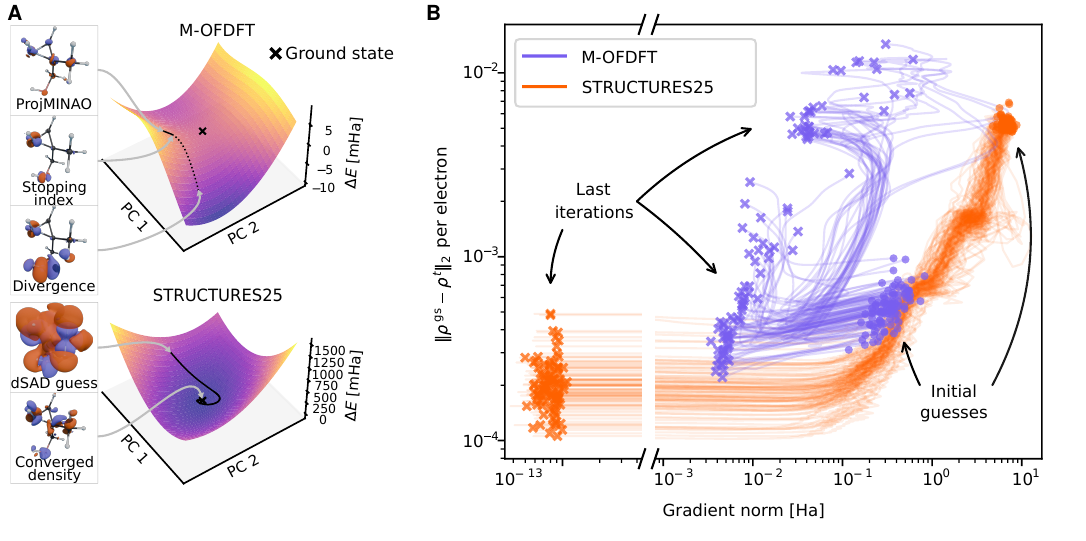}
    \caption{
    \textbf{STRUCTURES25 truly converges}.
    (\textbf{A}) Energy surfaces of the same QM9 molecule, according to the M-OFDFT and STRUCTURES25 functionals. Left: Density differences to ground state. The principal component analysis is performed on the respective density optimization trajectories. The M-OFDFT functional exhibits a saddle point and gradient descent diverges. The STRUCTURES25 functional has a minimum, which gradient descent with momentum finds even though starting from a cheaper, less accurate starting guess.
    (\textbf{B}) $L_2$ density error and gradient norm across density optimization on the same 100 random molecules from the QM9 test set. Both models ran for 6000 optimization iterations regardless of convergence.
    The respective initial guesses, projected MINAO for M-OFDFT ($\mathcal{O}(N^3)$), and dSAD ($\mathcal{O}(N)$) for STRUCTURES25, are marked by circles and last iterations are shown by crosses. The starting guess employed by M-OFDFT is an order of magnitude more accurate, but a considerable fraction of densities actually deteriorates across iterations. STRUCTURES25 almost monotonically improves densities across iterations, and gradient norms around $\text{10}^{-\text{13}}$ indicate proper convergence to a stable solution.}
    \label{fig:convergence}
\end{figure}

Ensuring that the learned energy functional has a true minimum at the correct ground state electron density (or very nearby) is of paramount practical importance (Fig.~\ref{fig:convergence}A). 
Indeed, it enables convergent variational density optimization to meaningful densities (Fig.~\ref{fig:convergence}B).
If no such minimum were present, following the gradient on such a misleading surface would result in unphysical densities, e.g., allowing electrons to collapse into a nucleus. 
Previously, such malformed energy surfaces required elaborate procedures to salvage a prediction from a diverging density optimization trajectory~\cite{zhang2024overcoming}.
In addition, an estimated ground state density with vanishing gradients 
$\nabla_{\vp}E=0$ 
is vital for the calculation of valid nuclear gradients, which are required for robust geometry optimization but especially for faithful molecular dynamics.

Improving upon the varied data generation pioneered in KineticNet~\cite{remme2023kineticnet} and combining it with a generalization~\cite{lippmann2025beyond} of the M-OFDFT architecture~\cite{zhang2024overcoming} finally affords a well-formed energy functional, which we call STRUCTURES25. 
Indeed, density optimization using this functional dramatically improves convergence on the QM9~\cite{ruddigkeit2012enumeration, ramakrishnan2014quantum} test set %
while reducing energy and density errors at the same time. 

An important side effect of these improvements are reduced demands on the quality, and hence cost, of the initial electron density for the OF-DFT calculation.
In fact, the robustness of the new functional allows starting from our version of a simple but exceedingly fast data-driven superposition of atomic densities (dSAD) guess, see materials and methods. %

In the following, we outline the changes to training data generation and architecture that made these improvements possible. 

\section*{Generating diverse training data}
Data efficiency in machine learning can be increased by using appropriate inductive biases, such as baking permutation or rotation equivariance into the architecture, or invoking prior knowledge such as scaling laws~\cite{parr1989density, hollingsworth2018can}, or more broadly learning an objective function rather than its minimizers. %
But even then, broad coverage of conceivable inputs in the training data is required. 

\begin{figure}[p]
    \centering
    \includegraphics[width=\linewidth]{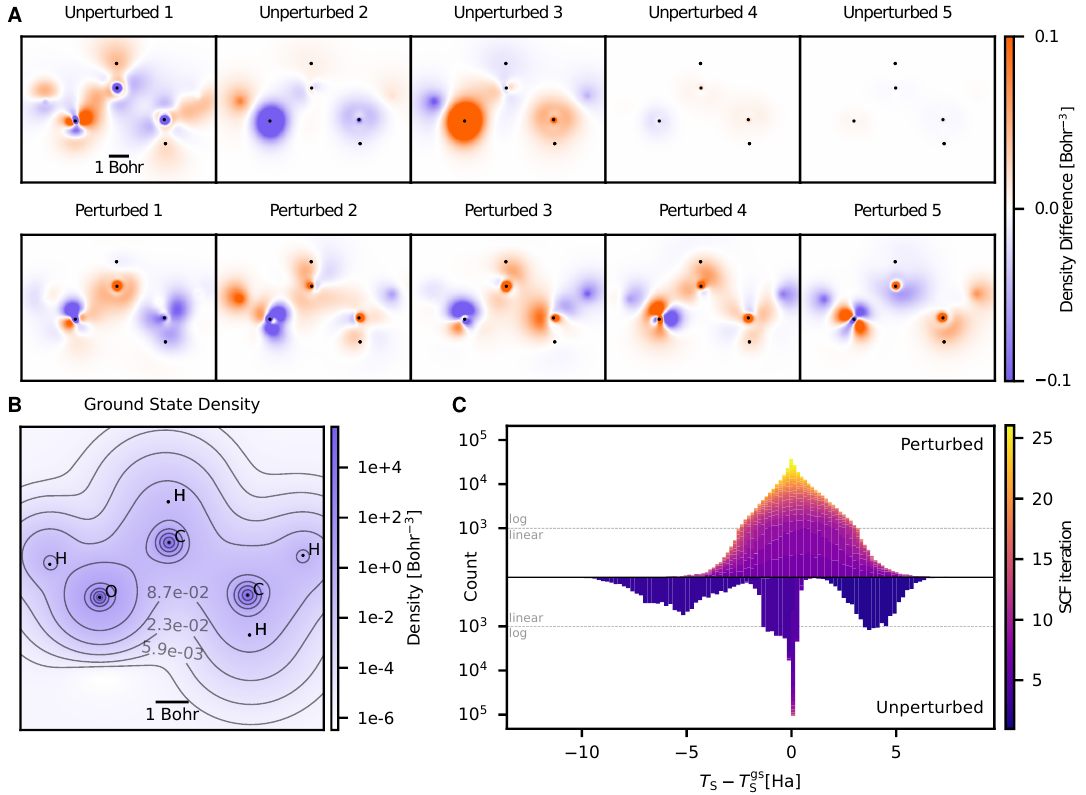}
    \caption{
    \textbf{Perturbation of the effective potential produces varied training data.}
    (\textbf{A}) Difference between various label densities and the ground state density, illustrating the proposed diversified training data. The unperturbed iterations (first row) demonstrate the rapid convergence of the standard Kohn-Sham procedure, with minimal changes observed from iteration 3 onwards. For the perturbed ones (second row), we intentionally perturbed the Fock matrix to disrupt convergence, resulting in increased diversity in the resulting electron densities.
    (\textbf{B}) Ground state electron density of an ethanol molecule, sliced through its symmetry axis. 
    (\textbf{C}) Histogram of the difference between the non-interacting kinetic energy \(T_\mathrm{S}\) of each sample and the corresponding ground state non-interacting kinetic energy \(T_\mathrm{S}^{\mathrm{gs}}\) (note the symlog-scale). Perturbing the effective potential \(V_{\mathrm{eff}}\) leads to a much more balanced distribution of \(T_\mathrm{S}\) around the value at the ground state \(T_\mathrm{S}^{\mathrm{gs}}\).}
    \label{fig:datagen}
\end{figure}

Zhang et al.~\cite{zhang2024overcoming} have used the insight that each iteration of the self-consistent Kohn Sham DFT procedure~\cite{kohn1965self}
\begin{equation} \label{eq:kohn_sham_equations_perturbed}
\left[-\frac{1}{2}\nabla^2 + V_\mathrm{eff}[\{\phi_i^{t-1}\}]  \right] \phi^t_i = \varepsilon^t_i \phi^t_i
\end{equation}
yields a consistent tuple of potential $V_\mathrm{eff}$, orbitals $\phi_i$ and associated energies $\varepsilon_i$ from which a training sample (density coefficients $\vp$, energy $E$, gradient $\nabla_{\vp}E$) can be obtained.  
The bottom of Fig.~\ref{fig:datagen}C characterizes the resulting high-dimensional training data in terms of a single axis, the energy difference between training and respective ground state electron densities. 
It is in the nature of self-consistent field (SCF) iterations that these quickly converge to the vicinity of the ground state, resulting in an uneven training distribution that is mostly concentrated in a spike around the ground state density.

We instead modify the potential in the above equation to read $ V_\mathrm{eff}[\{\phi_i^{t-1}\}] + \Delta^t$, where \mbox{$\Delta^t\colon \mathbb{R}^3\to\mathbb{R}$} are randomly sampled perturbations. 
This approach results in more varied training labels (Figs.~\ref{fig:datagen}A and \ref{fig:datagen}B) which in turn afford the training of much better-behaved functionals, see Fig.~\ref{fig:convergence} and ablation experiments in  supporting text \ref{sec:ablations}.

\section*{Training targets}
The question of which target to train against is a highly interesting one. 
According to Hohenberg and Kohn~\cite{hohenberg1964inhomogeneous}, the total electronic energy can be decomposed into a universal (independent of the external potential) functional $F[\rho]$ and a classical electrostatic interaction between the electron density and the external potential representing the atomic nuclei:  $E[\rho]= F[\rho] + \int \mathrm d^3 \mathbf r\, v_\mathrm{ext}(\mathbf r)\rho(\mathbf r)$.  
Following Levy and Lieb~\cite{parr1989density}, it is customary to further decompose the universal functional into a ``non-interacting kinetic energy'' $T_\mathrm{S}$, a classical electrostatic electron-electron or ``Hartree'' interaction $E_\mathrm{H}$ and an ``exchange-correlation'' term $E_\mathrm{XC}$, %
which captures all quantum effects not accounted for elsewhere.
The immense practical success of Kohn-Sham DFT is owing to the fact that reasonably good approximations to the true $E_\mathrm{XC}$ have been found~\cite{lee1988development,becke1993density,furness2020accurate}, with active efforts~\cite{kirkpatrick2021pushing} underway to further improve upon these using machine learning.

At first sight, learning a density functional form of just $T_\mathrm{S}$, the raison d'\^{e}tre of Kohn-Sham DFT, seems natural in the spirit of reductionism.
As an example, the model from M-OFDFT~\cite{zhang2024overcoming} evaluated in  Fig.~\ref{fig:overview}A was trained on the difference of this non-interacting kinetic energy and the classical APBEK functional~\cite{constantin2011semiclassical} in the fashion of ``delta learning.''

Here we instead opt to learn the sum of kinetic and exchange correlation energies because that eliminates the need for a quadrature grid which is otherwise required for the evaluation of most exchange correlation functionals. 
This target, denoted $E_\mathrm{TXC}$ in the following, aggregates all contributions that are not known analytically and gains further justification from the conjointness conjecture~\cite{constantin2011semiclassical}.

Finally, to obtain well-formed energy surfaces, it helps to use not only energies but also their functional derivatives as training targets. In the finite basis, the functional derivatives are gradients $\nabla_{\vp}E$. Obtaining these is not trivial, and details can be found in the supporting information, section \ref{sec:data_gen_appendix}.

\section*{Representation and Architecture}

A compact representation of electron density $\rho$ can be obtained in terms of a linear combination \cite{grisafi2018transferable, vergara2023efficient} of atom-centered basis functions $\{\omega_{\mu}\}$,~
\begin{equation}\label{eq:density_basis}
    \rho(\mathbf r) = \sum_\mu p_\mu \omega_\mu(\mathbf r)\,.
\end{equation}
We use an even-tempered Gaussian basis~\cite{bardo1974eventempered} $\{\omega_\mu\}$. 
While this representation does not guarantee positivity, regions of unphysical negative densities are no failure case that we encounter in practice, see Fig.~\ref{fig:negative_density_comparison} in the supporting information.

The presently most successful class of architectures in molecular machine learning are atomistic message passing graph neural networks~\cite{duval2023hitchhiker}. 
These iteratively exchange messages between atoms, along edges which are typically not defined by chemical bonds but rather by some distance cutoff or even by a fully connected graph.
As desired, atomistic message passing predictions of molecular properties are invariant to the (mostly arbitrary) order in which the constituent atoms are presented.
Similarly, as physical quantities transform equivariantly when the entire system is translated or rotated, so should the predictions. Scalar quantities, such as the energy, should be $E(3)$-invariant.

This equivariance with respect to rigid motions is commonly accomplished either by relying on the tensor product~\cite{thomas2018tensor} as basic bilinear operation, or by ``local canonicalization''~\cite{luo2022equivariant,kaba2023equivariance,lippmann2025beyond}.
The latter finds local coordinate systems, equivariant ``local frames,'' for each atom based on its few nearest neighbors. 

Having experimented extensively with either class of architecture~\cite{liao2024equiformerv2,simeon2024tensornet, ying2021transformers}, we obtain broadly comparable results with representatives of both; but we currently find the best trade-off between cost and performance with a Graphormer~\cite{ying2021transformers,shi2022benchmarking} type architecture. The latter profits from a self-attention mechanism, but is limited by the fact that it can only send scalar messages between nodes. In response, by invoking a recently proposed formalism~\cite{lippmann2025beyond}, we generalize the architecture to allow sending tensorial messages between nodes.

The input to our model consists of the density coefficients $\vp$ from Eq.~\ref{eq:density_basis} as well as the molecular geometry $\mathcal{M}$ given by atom positions $\{\mathbf{R}_a\}$ and types $\{Z_a\}$. The model predicts the energy $E$ while its gradient $\nabla_{\mathbf{p}}E$ is computed variationally using automated differentiation.

The atomic basis functions $\{\omega_{\mu}\}$ overlap and so the density coefficients are not independent.
We follow M-OFDFT~\cite{zhang2024overcoming} 
in first transforming the coefficients into an orthonormal basis by means of a global ``natural reparametrization'' and then subjecting
coefficients and energy gradients to dimension-wise rescaling and atomic reference modules, see~\cite{zhang2024overcoming} for details.

Importantly, for larger molecules, we use a distance cutoff in the definition of atomic adjacency, making for a message passing mechanism that scales gracefully with system size. 
For detailed model specifications, see the supporting text.

\subsection*{Evaluating STRUCTURES25}

We concentrate here on equilibrium-geometry organic molecules with a mass of up to a few hundred Dalton. While they span only a tiny corner of the entire chemical space, this family is already estimated to comprise anywhere between $10^{24}$ and $10^{60}$ members~\cite{ertl2003cheminformatics,reymond2010chemical}. 
Needless to say, this number grows further when taking larger biopolymers such as peptides or nucleotide chains into account.%

\begin{figure}[p]
    \centering
    \includegraphics[width=\linewidth]{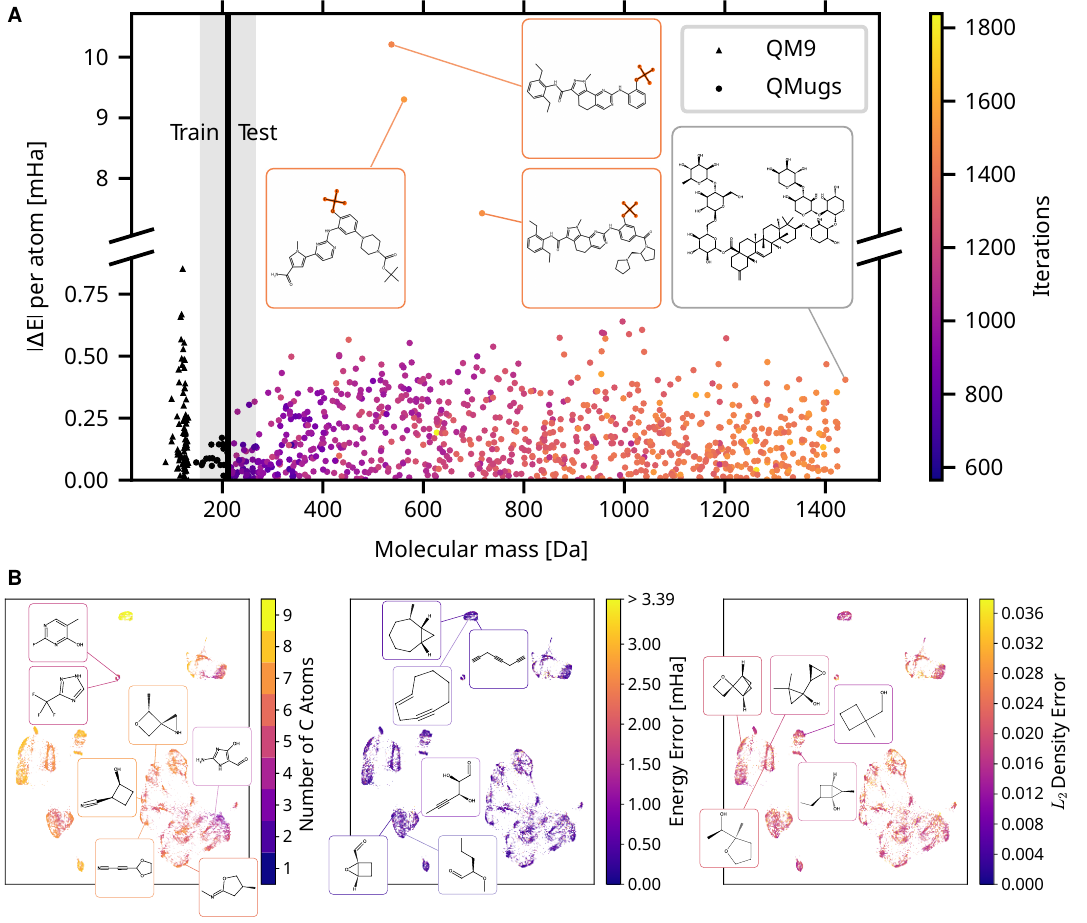}
    \caption{
    \textbf{STRUCTURES25 successfully extrapolates to larger molecules and shows vestiges of chemical ``understanding.''}
    (\textbf{A}) While the model is trained on molecules with strictly less than 16 heavy atoms, it still manages to generalize to larger molecules. 
    Three energy outliers (orange boxes) contain a trifluoromethoxy group, found only in a single molecule (with three conformers) in the training set. The gray box shows the largest molecule in our QMugs test set.
    (\textbf{B}) UMAP plot of internal activations before the first Graphormer layer. Left, middle and right:  Number of carbon atoms, energy error and density error after density fitting, respectively. 
    The groupings of related molecules in the plot may be interpreted as evidence for the emergence of a meaningful representation of chemistry in the model.}
    \label{fig:results}
\end{figure}

The QM9 database~\cite{ramakrishnan2014quantum}, while restricted to stable molecules and relaxed geometries, is already quite diverse when considering only small organic molecules, see Fig.~\ref{fig:results}B.
Orbital-free DFT as implemented by the STRUCTURES25 functional now affords density optimization which fully converges for each and every of the ca.~13k QM9 test molecules; with the resulting densities deviating from Kohn-Sham ground truth by $5.86\cdot10^{-4}$ in the metric of the squared error integrated over space, normalized by the number of electrons. 
The biggest contribution to this deviation comes not from the machine learning model, but stems from the intrinsic error of density fitting (see Fig.~\ref{fig:norm-triangle} in the supporting information), the process of expressing a Kohn-Sham density in the specific basis $\{\omega_{\mu}\}$ used to expand the orbital-free density in Eq.~\ref{eq:density_basis}. 
 Mean absolute energy errors of 0.64~mHa relative to the PBE/6-31G(2df,p) KS-DFT reference are well below the barrier of 1.6~mHa, a widely accepted definition of ``chemical accuracy''. 

Reaching Kohn-Sham accuracy and fully convergent density optimization on the full chemistry embodied by the QM9 dataset is already most auspicious for orbital-free DFT. Yet, for future applications, reliable extrapolation to larger systems is essential: It is here that standard Kohn-Sham DFT becomes intractable. 
Following prior work~\cite{zhang2024overcoming}, we evaluate extrapolation accuracy on the QMugs database~\cite{isert2022qmugs} comprising substantially larger druglike molecules. 
To this end, we train on smaller molecules with a mass up to around 200~Da, and evaluate on a subset of 850 test molecules, with a mass up to around 1400~Da. On these, the STRUCTURES25 functional achieves fully convergent density optimization on all molecules, shown in Fig.~\ref{fig:results}. Remarkably, even though the STRUCTURES25 functional uses message passing only across local neighborhoods up to six Bohr in radius, the mean absolute error per atom is larger, but does not grow with the size of the molecule, across the QMugs database (Fig.~\ref{fig:results}A). 
The three largest energy errors were all associated with the trifluoromethoxy group, a moiety which turned out to be represented only in terms of a single molecule in the training data. 
Larger molecules in the QMugs database highlight the improved scaling of machine-learned OF-DFT in comparison to the Kohn-Sham reference calculations, resulting in nearly an order of magnitude speedup on identical hardware (section \ref{sec:runtime_comparison} in the supporting information). While density optimization requires many more iterations than a typical Kohn-Sham SCF calculation, this is more than offset by the relative simplicity of each single iteration which does not require matrix diagonalization. 

\begin{table}[h!]
\renewcommand{\arraystretch}{1.2}
\begin{threeparttable}
\resizebox{\textwidth}{!}{\begin{minipage}{\textwidth}
    \centering
    \caption{
    \textbf{Orbital-free DFT finds ground state energies with chemical accuracy relative \\
    \noindent  to a PBE/6-31G(2df,p) Kohn-Sham reference, and meaningful densities on QM9 and QMugs.}}
    \label{tab:results}
    \footnotesize
    \begin{tabular}{lccccccccc} \thickhline
        Dataset & Functional & Local\tnote{a} & $|\Delta E|$\tnote{b} & $|\Delta E|/ N$\tnote{c} & $\| \Delta \rho \|_2$\tnote{d} & $\| \Delta \rho \|_{2} / N_e$\tnote{e}& Runtime\tnote{f}\\
        & & & (mHa) & (mHa) & ($\text{10}^{-\text{2}}$) & ($\text{10}^{-\text{4}}$)& (s)\\
        \hline
        QM9 & Ours & $\times$ &  $\mathbf{0.644 \boldsymbol\pm 0.006}$ & $\mathbf{0.0380 \boldsymbol\pm 0.0003}$ & $\mathbf{1.40 \boldsymbol\pm 0.02}$ & $\mathbf{2.12 \boldsymbol\pm 0.03}$ & \phantom{0}\textbf{13} \\
        & M-OFDFT\tnote{g} & $\times$ & 1.37 & 0.088 & 2.7 & 4.2 & 183\\
        \hline
        QMugs & Ours & $\checkmark$ & $25 \pm 2 $ & $0.231 \pm 0.019$  &  $\mathbf{6.8 \boldsymbol\pm 0.2}$ & $\mathbf{1.61 \boldsymbol \pm 0.04}$ & \phantom{0}\textbf{40}\\
        & M-OFDFT & $\times$ & \textbf{18} & \textbf{0.17} & 7.0 & 1.8 & 319 \\ \thickhline
    \end{tabular}
    \begin{tablenotes}
    \item[a] Indicates whether the model guarantees a finite field of view.
    \item[b, c] Mean absolute energy error, total and per atom.
    \item[d, e] Mean $L_2$ norm of the density difference, total and per electron.
    \item[f] Average time for density optimization for a single molecule on an Nvidia Quadro RTX 6000 GPU for QM9 and an Nvidia A100 GPU for QMUGS.
    \item[g] Trained on the $T_\mathrm{S}-\mathrm{APBEK}$ target (all other models predict $E_\mathrm{TXC}$). 
    
    \end{tablenotes}
\end{minipage}
}
\end{threeparttable}
\end{table}

\subsection*{Orbital-free DFT: Status quo, and where next}
Building on a compact representation of the electron density~\cite{grisafi2018transferable,vergara2023efficient} (Eq.~\ref{eq:density_basis}), state-of-the-art machine learning architectures~\cite{ying2021transformers,lippmann2025beyond}, automated differentiation~\cite{paszke2019pytorch}, efficient training data generation and reparametrization~\cite{zhang2024overcoming} and a strategy to create more balanced training samples~\cite{remme2023kineticnet}, orbital-free DFT is now emerging as a viable framework for accurate calculations grounded in the Hohenberg-Kohn theorems.

Still, important limitations persist. The present functional, trained on organic equilibrium-geometry molecules only, does not yet generalize to high-energy geometries (a prerequisite for successful geometry optimization), or across the periodic table. Also, the poor predictions of the trifluoromethoxy group illustrate that with the present size of training data, sufficient examples of chemical moietys must be seen during training. It will be fascinating to see to what extent future models, trained on more extensive data, will ``grok'' chemistry and generalize across more of chemical space.
Indeed, we expect future machine-learned density functionals to work well in those regimes in which Kohn-Sham DFT does.
In particular, the viable area should be significantly larger than the kind of chemistries typified by the QM9 and QMugs databases. 

For the following discussion, we single out the question of scaling, one principal reason to study OF-DFT in the first place.
To become an alternative to linear-scaling KS-DFT~\cite{soler2002siesta, hutter2014cp2k, bowler2006recent, skylaris2005introducing}, OF-DFT will not just need quasi-linear scaling, but also a small prefactor. %
Let us recall the computational complexity bottlenecks of OF-DFT, beginning with the initialization. 
While many classical OF-DFT functionals are sufficiently robust to start from cheap initial density guesses, the previous state of the art, the machine-learned M-OFDFT~\cite{zhang2024overcoming} was dependent on MINAO~\cite{almlf1982principle,vanlenthe2006superposition} which has cubic complexity in the number of atoms $N$. The robustness of  STRUCTURES25 allows starting from our simple dSAD guess (see supporting text) which scales linearly. The next major step in the computational pipeline is
natural reparametrization, which still has $\mathcal{O}(N^3)$ complexity. Finding cheaper alternatives has top priority. 
Next, message passing on completely connected graphs comes with $\mathcal{O}(N^2)$ complexity. Our local model operating on a radius graph reduces this to $\mathcal{O}(N)$. 
The cost of the Hartree term scales quadratically with the number of basis functions when the representation in Eq.~\ref{eq:density_basis} is evoked. 
Approximations such as the fast multipole method can bring this cost down to linearithmic scaling.

Outside orbital-free DFT, direct property prediction using ``foundation models'' is progressing with great strides\cite{batatia2024foundation, yuan2025foundationmodelsatomisticsimulation, isayev2025aimnet2, facebookUMA}. When striving to make valid predictions across a broad range of chemical space, a key question is which approach will prevail: A transductive approach, as in foundation models for property prediction which directly map from molecular geometry to observable; or an inductive ansatz, as in orbital-free DFT, where a universal functional is learned and variational optimization yields the electron density and its energy, from which observables can be derived. We believe the inductive approach will generalize better, but are eager to learn whether this intuition stands the test of time. 

Overall, orbital-free DFT is now on the cusp of becoming practically useful in the molecular realm, due in part to the present contributions building on excellent prior work~\cite{chan01thomas-fermi-dirac,
ying2021transformers, remme2023kineticnet,zhang2024overcoming,lippmann2025beyond}.

\clearpage %

\bibliography{structures25}

\begin{thebibliography}{10}
\providecommand{\url}[1]{\texttt{#1}}
\expandafter\ifx\csname urlstyle\endcsname\relax
  \providecommand{\doi}[1]{doi:\discretionary{}{}{}#1}\else
  \providecommand{\doi}{doi:\discretionary{}{}{}\begingroup \urlstyle{rm}\Url}\fi

\bibitem{hohenberg1964inhomogeneous}
P.~Hohenberg, W.~Kohn, Inhomogeneous electron gas. \emph{Phys. Rev.} \textbf{136}~(3B), B864 (1964).

\bibitem{weizsacker1935theorie}
C.~v. Weizs{\"a}cker, Zur Theorie der Kernmassen. \emph{Z. Phys.} \textbf{96}~(7), 431--458 (1935).

\bibitem{thomas1927calculation}
L.~H. Thomas, The calculation of atomic fields, in \emph{Mathematical proceedings of the Cambridge philosophical society} (Cambridge University Press), vol.~23 (1927), pp. 542--548.

\bibitem{fermi1928statistische}
E.~Fermi, {Eine statistische Methode zur Bestimmung einiger Eigenschaften des Atoms und ihre Anwendung auf die Theorie des periodischen Systems der Elemente}. \emph{Z. Phys.} \textbf{48}~(1), 73--79 (1928).

\bibitem{kohn1965self}
W.~Kohn, L.~J. Sham, Self-{Consistent} {Equations} {Including} {Exchange} and {Correlation} {Effects}. \emph{Phys. Rev.} \textbf{140}~(4A), A1133--A1138 (1965).

\bibitem{kohn1996density}
W.~Kohn, Density functional and density matrix method scaling linearly with the number of atoms. \emph{Phys. Rev. Lett.} \textbf{76}~(17), 3168 (1996).

\bibitem{yang1986molecules}
W.~Yang, R.~G. Parr, C.~Lee, Various functionals for the kinetic energy density of an atom or molecule. \emph{Phys. Rev. A} \textbf{34}, 4586--4590 (1986), \doi{10.1103/PhysRevA.34.4586}, \url{https://link.aps.org/doi/10.1103/PhysRevA.34.4586}.

\bibitem{wang1992nonlocal}
L.-W. Wang, M.~P. Teter, Kinetic-energy functional of the electron density. \emph{Physical review. B, Condensed matter} \textbf{45}~(23), 13196--13220 (1992).

\bibitem{lembarki1994fromxc}
A.~Lembarki, H.~Chermette, Obtaining a gradient-corrected kinetic-energy functional from the Perdew-Wang exchange functional. \emph{Phys. Rev. A} \textbf{50}, 5328--5331 (1994), \doi{10.1103/PhysRevA.50.5328}, \url{https://link.aps.org/doi/10.1103/PhysRevA.50.5328}.

\bibitem{thakkar1992comparison}
A.~J. Thakkar, Comparison of kinetic-energy density functionals. \emph{Phys. Rev. A} \textbf{46}~(11), 6920 (1992).

\bibitem{wang1999orbital}
Y.~A. Wang, N.~Govind, E.~A. Carter, Orbital-free kinetic-energy density functionals with a density-dependent kernel. \emph{Phys. Rev. B} \textbf{60}~(24), 16350 (1999).

\bibitem{constantin2011semiclassical}
L.~A. Constantin, E.~Fabiano, S.~Laricchia, F.~Della~Sala, Semiclassical neutral atom as a reference system in density functional theory. \emph{Phys. Rev. Lett.} \textbf{106}~(18), 186406 (2011).

\bibitem{karasiev2013nonempirical}
V.~V. Karasiev, D.~Chakraborty, O.~A. Shukruto, S.~Trickey, Nonempirical generalized gradient approximation free-energy functional for orbital-free simulations. \emph{Physical Review B—Condensed Matter and Materials Physics} \textbf{88}~(16), 161108 (2013).

\bibitem{luo2018simple}
K.~Luo, V.~V. Karasiev, S.~Trickey, A simple generalized gradient approximation for the noninteracting kinetic energy density functional. \emph{Physical Review B} \textbf{98}~(4), 041111 (2018).

\bibitem{Witt_delRio_Dieterich_Carter_2018}
W.~C. Witt, B.~G. del Rio, J.~M. Dieterich, E.~A. Carter, Orbital-free density functional theory for materials research. \emph{Journal of Materials Research} \textbf{33}~(7), 777–795 (2018), \doi{10.1557/jmr.2017.462}.

\bibitem{shao2021revised}
X.~Shao, W.~Mi, M.~Pavanello, Revised Huang-Carter nonlocal kinetic energy functional for semiconductors and their surfaces. \emph{Physical Review B} \textbf{104}~(4), 045118 (2021).

\bibitem{mi2023orbital}
W.~Mi, K.~Luo, S.~Trickey, M.~Pavanello, Orbital-free density functional theory: An attractive electronic structure method for large-scale first-principles simulations. \emph{Chem. Rev.} \textbf{123}~(21), 12039--12104 (2023).

\bibitem{snyder2012finding}
J.~C. Snyder, M.~Rupp, K.~Hansen, K.-R. M{\"u}ller, K.~Burke, Finding density functionals with machine learning. \emph{Phys. Rev. Lett.} \textbf{108}~(25), 253002 (2012).

\bibitem{kulik2022roadmap}
H.~J. Kulik, \emph{et~al.}, Roadmap on machine learning in electronic structure. \emph{Electronic Structure} \textbf{4}~(2), 023004 (2022).

\bibitem{bogojeski2020quantum}
M.~Bogojeski, L.~Vogt-Maranto, M.~E. Tuckerman, K.-R. M{\"u}ller, K.~Burke, Quantum chemical accuracy from density functional approximations via machine learning. \emph{Nature communications} \textbf{11}~(1), 5223 (2020).

\bibitem{yao2016kinetic}
K.~Yao, J.~Parkhill, Kinetic Energy of Hydrocarbons as a Function of Electron Density and Convolutional Neural Networks. \emph{J. Chem. Theory Comput.} \textbf{12}~(3), 1139--1147 (2016).

\bibitem{seino2018semi-local}
J.~Seino, R.~Kageyama, M.~Fujinami, Y.~Ikabata, H.~Nakai, Semi-local machine-learned kinetic energy density functional with third-order gradients of electron density \textbf{148}~(24), 241705.

\bibitem{fujinami2020orbital-free}
M.~Fujinami, R.~Kageyama, J.~Seino, Y.~Ikabata, H.~Nakai, Orbital-free density functional theory calculation applying semi-local machine-learned kinetic energy density functional and kinetic potential. \emph{Chem. Phys. Lett.} \textbf{748}, 137358 (2020).

\bibitem{ryczko2022toward}
K.~Ryczko, S.~J. Wetzel, R.~G. Melko, I.~Tamblyn, Toward Orbital-Free Density Functional Theory with Small Data Sets and Deep Learning \textbf{18}~(2), 1122--1128 (2022), publisher: American Chemical Society.

\bibitem{remme2023kineticnet}
R.~Remme, T.~Kaczun, M.~Scheurer, A.~Dreuw, F.~A. Hamprecht, {KineticNet}: Deep learning a transferable kinetic energy functional for orbital-free density functional theory. \emph{J. Chem. Phys.} \textbf{159}~(14), 144113 (2023).

\bibitem{zhang2024overcoming}
H.~Zhang, \emph{et~al.}, Overcoming the barrier of orbital-free density functional theory for molecular systems using deep learning. \emph{Nat. Comput. Sci.} \textbf{4}, 210--223 (2024).

\bibitem{grisafi2018transferable}
A.~Grisafi, \emph{et~al.}, Transferable machine-learning model of the electron density. \emph{ACS Central Science} \textbf{5}~(1), 57--64 (2018).

\bibitem{vergara2023efficient}
U.~A. Vergara-Beltran, J.~I. Rodr{\'\i}guez, An efficient zero-order evolutionary method for solving the orbital-free density functional theory problem by direct minimization. \emph{J. Chem. Phys.} \textbf{159}~(12), 124102 (2023).

\bibitem{ruddigkeit2012enumeration}
L.~Ruddigkeit, R.~Van~Deursen, L.~C. Blum, J.-L. Reymond, Enumeration of 166 billion organic small molecules in the chemical universe database GDB-17. \emph{J. Chem. Inf. Model.} \textbf{52}~(11), 2864--2875 (2012).

\bibitem{ramakrishnan2014quantum}
R.~Ramakrishnan, P.~O. Dral, M.~Rupp, O.~A. Von~Lilienfeld, Quantum chemistry structures and properties of 134 kilo molecules. \emph{Sci. Data} \textbf{1}~(1), 140022 (2014).

\bibitem{lippmann2025beyond}
P.~Lippmann, G.~Gerhartz, R.~Remme, F.~A. Hamprecht, Beyond Canonicalization: How Tensorial Messages Improve Equivariant Message Passing, in \emph{The Thirteenth International Conference on Learning Representations} (2025).

\bibitem{parr1989density}
R.~Parr, W.~Yang, \emph{Density-Functional Theory of Atoms and Molecules} (Oxford University Press) (1989).

\bibitem{hollingsworth2018can}
J.~Hollingsworth, L.~Li, T.~E. Baker, K.~Burke, Can exact conditions improve machine-learned density functionals? \emph{J. Chem. Phys.} \textbf{148}~(24), 241743 (2018).

\bibitem{lee1988development}
C.~Lee, W.~Yang, R.~G. Parr, Development of the Colle-Salvetti correlation-energy formula into a functional of the electron density. \emph{Phys. Rev. B} \textbf{37}~(2), 785 (1988).

\bibitem{becke1993density}
A.~D. Becke, Density‐functional thermochemistry. III. The role of exact exchange. \emph{J. Chem. Phys.} \textbf{98}~(7), 5648--5652 (1993).

\bibitem{furness2020accurate}
J.~W. Furness, A.~D. Kaplan, J.~Ning, J.~P. Perdew, J.~Sun, Accurate and Numerically Efficient r2SCAN Meta-Generalized Gradient Approximation. \emph{J. Phys. Chem. Lett.} \textbf{11}~(19), 8208--8215 (2020).

\bibitem{kirkpatrick2021pushing}
J.~Kirkpatrick, \emph{et~al.}, Pushing the frontiers of density functionals by solving the fractional electron problem. \emph{Science} \textbf{374}~(6573), 1385--1389 (2021).

\bibitem{bardo1974eventempered}
R.~D. Bardo, K.~Ruedenberg, {Even‐tempered atomic orbitals. VI. Optimal orbital exponents and optimal contractions of Gaussian primitives for hydrogen, carbon, and oxygen in molecules}. \emph{J. Chem. Phys.} \textbf{60}~(3), 918--931 (1974).

\bibitem{duval2023hitchhiker}
A.~Duval, \emph{et~al.}, A Hitchhiker's Guide to Geometric GNNs for 3D Atomic Systems. \emph{arXiv Preprint}  (2024), arXiv:2312.07511 [cs.LG].

\bibitem{thomas2018tensor}
N.~Thomas, \emph{et~al.}, Tensor field networks: Rotation- and translation-equivariant neural networks for 3D point clouds. \emph{arXiv Preprint}  (2018), arXiv:1802.08219 [cs.LG].

\bibitem{luo2022equivariant}
S.~Luo, \emph{et~al.}, Equivariant point cloud analysis via learning orientations for message passing, in \emph{Proceedings of the IEEE/CVF Conference on Computer Vision and Pattern Recognition} (2022), pp. 18932--18941.

\bibitem{kaba2023equivariance}
S.-O. Kaba, A.~K. Mondal, Y.~Zhang, Y.~Bengio, S.~Ravanbakhsh, Equivariance with learned canonicalization functions, in \emph{International Conference on Machine Learning} (PMLR) (2023), pp. 15546--15566.

\bibitem{liao2024equiformerv2}
Y.-L. Liao, B.~M. Wood, A.~Das, T.~Smidt, EquiformerV2: Improved Equivariant Transformer for Scaling to Higher-Degree Representations, in \emph{The Twelfth International Conference on Learning Representations} (2024).

\bibitem{simeon2024tensornet}
G.~Simeon, G.~De~Fabritiis, {TensorNet}: Cartesian Tensor Representations for Efficient Learning of Molecular Potentials, in \emph{Advances in Neural Information Processing Systems}, vol.~36 (2023), pp. 37334--37353.

\bibitem{ying2021transformers}
C.~Ying, \emph{et~al.}, Do transformers really perform badly for graph representation?, in \emph{Advances in neural information processing systems}, vol.~34 (2021), pp. 28877--28888.

\bibitem{shi2022benchmarking}
Y.~Shi, \emph{et~al.}, Benchmarking Graphormer on Large-Scale Molecular Modeling Datasets. \emph{arXiv Preprint}  (2023), arXiv:2203.04810 [cs.LG].

\bibitem{ertl2003cheminformatics}
P.~Ertl, Cheminformatics analysis of organic substituents: identification of the most common substituents, calculation of substituent properties, and automatic identification of drug-like bioisosteric groups. \emph{J. Chem. Inf. Comput. Sci.} \textbf{43}~(2), 374--380 (2003).

\bibitem{reymond2010chemical}
J.-L. Reymond, R.~Van~Deursen, L.~C. Blum, L.~Ruddigkeit, Chemical space as a source for new drugs. \emph{Med. Chem. Comm.} \textbf{1}~(1), 30--38 (2010).

\bibitem{isert2022qmugs}
C.~Isert, K.~Atz, J.~Jim{\'e}nez-Luna, G.~Schneider, {QMugs}, quantum mechanical properties of drug-like molecules. \emph{Sci. Data} \textbf{9}~(1), 273 (2022).

\bibitem{paszke2019pytorch}
A.~Paszke, \emph{et~al.}, Pytorch: An imperative style, high-performance deep learning library. \emph{Advances in neural information processing systems} \textbf{32}, 8024--8035 (2019).

\bibitem{soler2002siesta}
J.~M. Soler, \emph{et~al.}, The {SIESTA} method for ab initio order-N materials simulation. \emph{J. Condens. Matter Phys.} \textbf{14}~(11), 2745 (2002).

\bibitem{hutter2014cp2k}
J.~Hutter, M.~Iannuzzi, F.~Schiffmann, J.~VandeVondele, cp2k: atomistic simulations of condensed matter systems. \emph{Wiley Interdiscip. Rev. Comput. Mol. Sci.} \textbf{4}~(1), 15--25 (2014).

\bibitem{bowler2006recent}
D.~Bowler, R.~Choudhury, M.~Gillan, T.~Miyazaki, Recent progress with large-scale ab initio calculations: the CONQUEST code. \emph{Phys. Status Solidi B} \textbf{243}~(5), 989--1000 (2006).

\bibitem{skylaris2005introducing}
C.-K. Skylaris, P.~D. Haynes, A.~A. Mostofi, M.~C. Payne, Introducing ONETEP: Linear-scaling density functional simulations on parallel computers. \emph{J. Chem. Phys.} \textbf{122}~(8), 084119 (2005).

\bibitem{almlf1982principle}
J.~Alml\"{o}f, K.~Faegri, K.~Korsell, Principles for a direct SCF approach to LICAO–MO ab‐initio calculations. \emph{J. Comput. Chem.} \textbf{3}~(3), 385–399 (1982).

\bibitem{vanlenthe2006superposition}
J.~H. Van~Lenthe, R.~Zwaans, H.~J.~J. Van~Dam, M.~F. Guest, Starting SCF calculations by superposition of atomic densities. \emph{J. Comput. Chem.} \textbf{27}~(8), 926–932 (2006).

\bibitem{batatia2024foundation}
I.~Batatia, \emph{et~al.}, A foundation model for atomistic materials chemistry. \emph{arXiv Preprint}  (2024), arXiv:2401.00096 [physics.chem-ph].

\bibitem{yuan2025foundationmodelsatomisticsimulation}
E.~C.~Y. Yuan, \emph{et~al.}, Foundation Models for Atomistic Simulation of Chemistry and Materials. \emph{arXiv Preprint}  (2025), arXiv:2503.10538 [physics.chem-ph], \url{https://arxiv.org/abs/2503.10538}.

\bibitem{isayev2025aimnet2}
O.~Isayev, D.~Anstine, R.~Zubaiuk, AIMNet2: a neural network potential to meet your neutral, charged, organic, and elemental-organic needs. \emph{Chemical Science} \textbf{16}, 10228--10244 (2025).

\bibitem{facebookUMA}
{Facebook AI}, Universal Multimodal Agent (UMA), \url{https://huggingface.co/facebook/UMA} (2025), accessed June 12, 2025.

\bibitem{chan01thomas-fermi-dirac}
G.~K.-L. Chan, A.~J. Cohen, N.~C. Handy, Thomas–Fermi–Dirac–von Weizsäcker models in finite systems. \emph{J. Chem. Phys.} \textbf{114}~(2), 631--638 (2001).

\bibitem{sun2020recent}
Q.~Sun, \emph{et~al.}, Recent developments in the {PySCF} program package. \emph{J. Chem. Phys.} \textbf{153}~(2) (2020).

\bibitem{sun2018pyscf}
Q.~Sun, \emph{et~al.}, PySCF: the Python-based simulations of chemistry framework. \emph{Wiley Interdiscip. Rev. Comput. Mol. Sci.} \textbf{8}~(1), e1340 (2018).

\bibitem{Sun2015}
Q.~Sun, Libcint: An efficient general integral library for Gaussian basis functions. \emph{J. Comput. Chem.} \textbf{36}~(22), 1664–1671 (2015).

\bibitem{krishnan1980a}
R.~Krishnan, J.~S. Binkley, R.~Seeger, J.~A. Pople, Self-consistent molecular orbital methods. XX. A basis set for correlated wave functions. \emph{J. Chem. Phys.} \textbf{72}, 650--654 (1980).

\bibitem{perdew1996generalized}
J.~P. Perdew, K.~Burke, M.~Ernzerhof, Generalized Gradient Approximation Made Simple. \emph{Phys. Rev. Lett.} \textbf{77}~(18), 3865--3868 (1996).

\bibitem{pulay1982improved}
P.~Pulay, Improved SCF convergence acceleration. \emph{J. Comput. Chem.} \textbf{3}~(4), 556--560 (1982).

\bibitem{kudin2002black}
K.~N. Kudin, G.~E. Scuseria, E.~Cancès, A black-box self-consistent field convergence algorithm: One step closer. \emph{J. Chem. Phys.} \textbf{116}~(19), 8255--8261 (2002).

\bibitem{whitten1973coulombic}
J.~L. Whitten, Coulombic potential energy integrals and approximations. \emph{J. Chem. Phys.} \textbf{58}~(10), 4496--4501 (1973).

\bibitem{dunlap1979some}
B.~I. Dunlap, J.~Connolly, J.~Sabin, On some approximations in applications of {X$\alpha$} theory. \emph{J. Chem. Phys.} \textbf{71}~(8), 3396--3402 (1979).

\bibitem{vahtras1993integral}
O.~Vahtras, J.~Alml{\"o}f, M.~Feyereisen, Integral approximations for LCAO-SCF calculations. \emph{Chem. Phys. Lett.} \textbf{213}~(5-6), 514--518 (1993).

\bibitem{ryczko2022orbitalfree}
K.~Ryczko, S.~J. Wetzel, R.~G. Melko, I.~Tamblyn, Toward {{Orbital-Free Density Functional Theory}} with {{Small Data Sets}} and {{Deep Learning}}. \emph{J. Chem. Theory Comput.} \textbf{18}~(2), 1122--1128 (2022).

\bibitem{shi2021inverse}
Y.~Shi, A.~Wasserman, Inverse {Kohn–Sham} Density Functional Theory: Progress and Challenges. \emph{J. Phys. Chem. Lett.} \textbf{12}~(22), 5308--5318 (2021).

\bibitem{batatia2022mace}
I.~Batatia, D.~P. Kovacs, G.~Simm, C.~Ortner, G.~Csanyi, {MACE}: Higher Order Equivariant Message Passing Neural Networks for Fast and Accurate Force Fields, in \emph{Advances in Neural Information Processing Systems}, vol.~35 (2022), pp. 11423--11436.

\bibitem{loewdin1950nonorthogonal}
P.-O. L{\"o}wdin, On the Non-Orthogonality Problem Connected with the Use of Atomic Wave Functions in the Theory of Molecules and Crystals. \emph{J. Chem. Phys.} \textbf{18}~(3), 365--375 (1950).

\bibitem{loewdin1970nonorthogonal}
P.-O. L{\"o}wdin, On the Nonorthogonality Problem, in \emph{Advances in {{Quantum Chemistry}}} (Elsevier), vol.~5, pp. 185--199 (1970).

\bibitem{aiken1980onLoewdin}
J.~G. Aiken, J.~A. Erdos, J.~A. Goldstein, {On L{\"o}wdin orthogonalization}. \emph{Int. J. Quant. Chem.} \textbf{18}~(4), 1101--1108 (1980).

\bibitem{aiken1975Loewdinminimum}
J.~G. Aiken, H.~B. Jonassen, H.~S. Aldrich, L{\"o}wdin Orthogonalization as a Minimum Energy Perturbation. \emph{J. Chem. Phys.} \textbf{62}~(7), 2745--2746 (1975).

\bibitem{loshchilov2016sgdr}
I.~Loshchilov, F.~Hutter, SGDR: Stochastic Gradient Descent with Warm Restarts, arxiv:1608.03983 [cs.LG] (2017).

\bibitem{loshchilov2018decoupled}
I.~Loshchilov, F.~Hutter, Decoupled Weight Decay Regularization, in \emph{International Conference on Learning Representations} (2019).

\bibitem{lehtola2019initialguess}
S.~Lehtola, Assessment of Initial Guesses for Self-Consistent Field Calculations. Superposition of Atomic Potentials: Simple yet Efficient. \emph{J. Chem. Theory Comput.} \textbf{15}~(3), 1593–1604 (2019).

\bibitem{weigend2005a}
F.~Weigend, R.~Ahlrichs, Balanced basis sets of split valence, triple zeta valence and quadruple zeta valence quality for H to Rn: Design and assessment of accuracy. \emph{Phys. Chem. Chem. Phys.} \textbf{7}, 3297 (2005).

\bibitem{boronski2005}
J.~Boroński, R.~M. Nieminen, Accurate exchange and correlation potentials for the electron gas. \emph{Phys. Rev. B} \textbf{72}~(12), 125115 (2005).

\bibitem{Becke1997}
A.~D. Becke, Density-functional thermochemistry. V. Systematic optimization of exchange-correlation functionals. \emph{J. Chem. Phys.} \textbf{107}, 8554--8560 (1997).

\bibitem{Adamo1999}
C.~Adamo, V.~Barone, Toward reliable density functional methods without adjustable parameters: The {PBE0} model. \emph{J. Chem. Phys.} \textbf{110}~(13), 6158--6170 (1999).

\bibitem{Lee1988}
C.~Lee, W.~Yang, R.~G. Parr, Development of the Colle-Salvetti correlation-energy formula into a functional of the electron density. \emph{Phys. Rev. B} \textbf{37}~(2), 785--789 (1988).

\bibitem{Perdew1986}
J.~P. Perdew, Y.~Wang, Accurate and simple analytic representation of the electron‐gas correlation energy. \emph{Phys. Rev. B} \textbf{33}, 8822--8824 (1986).

\bibitem{Becke1996}
A.~D. Becke, Density‐functional thermochemistry. IV. A new dynamical correlation functional and implications for exact exchange. \emph{J. Chem. Phys.} \textbf{104}~(7), 1040--1046 (1996).

\bibitem{li2024introducting}
R.~Li, Q.~Sun, X.~Zhang, G.~K.-L. Chan, Introducing GPU-acceleration into the Python-based Simulations of Chemistry Framework (2024), \url{https://arxiv.org/abs/2407.09700}.

\bibitem{wu2024enhancing}
X.~Wu, \emph{et~al.}, Enhancing GPU-acceleration in the Python-based Simulations of Chemistry Framework (2024), \url{https://arxiv.org/abs/2404.09452}.

\end{thebibliography}
\bibliographystyle{sciencemag}

\section*{Acknowledgments}
The authors thank the STRUCTURES excellence cluster for nurturing a unique collaborative environment that made this work possible in the first place. The authors also thank Maurits Haverkort, Corinna Steffen, Virginia Lenk and Andreas Hermann for helpful discussions, chemical insight and analytics.

\paragraph*{Funding:}
This work is supported by Deutsche Forschungsgemeinschaft (DFG) under Germany’s Excellence Strategy EXC-2181/1--390900948 (the Heidelberg STRUCTURES Excellence Cluster). 
The authors acknowledge support by the state of Baden-Württemberg through bwHPC
and the German Research Foundation (DFG) through grant INST 35/1597-1 FUGG.
C.A.G.~and M.V.K.\ were supported by the Konrad Zuse School of Excellence in Learning and Intelligent Systems (ELIZA) through the DAAD program Konrad Zuse Schools of Excellence in Artificial Intelligence, sponsored by the Federal Ministry of Education and Research. 
F.A.H.~acknowledges partial support by the Royal Society -- International Exchanges, IES\textbackslash R2\textbackslash 242159. 
T.K.~and A.D.~acknowledge support by the German Research Foundation (DFG) through grant no INST40/575-1 FUGG (JUSTUS 2 cluster). The authors also acknowledge partial funding by Deutsche Forschungsgemeinschaft (DFG, German Research Foundation) Projektnummber 281029004 -- SFB 1249 and Projektnummer 240245660 -- SFB 1129. 
We also acknowledge partial support by the Klaus Tschira Stiftung gGmbH in the framework of the SIMPLAIX consortium, as well as by the Carl-Zeiss-Stiftung. 

\paragraph*{Author contributions:}
Conceptualization: R.R., T.K., A.D., F.A.H.; Data Curation: R.R., T.K., M.K.I., M.V.K.; Funding acquisition: A.D., F.A.H.; Investigation: R.R., T.K., T.E., C.A.G., D.G., G.G., M.K.I., M.V.K., P.L., S.W., A.D., F.A.H.; Methodology: R.R., T.K., C.A.G., M.K.I., M.V.K., P.L., S.W., A.D., F.A.H.; Project Administration: R.R., F.A.H.; Software: R.R., T.K., T.E., C.A.G., D.G., G.G., M.K.I., M.V.K., P.L., J.S.S., S.W.; Supervision: A.D., F.A.H.; Visualization: R.R., T.K., T.E., C.A.G., D.G., G.G., M.K.I., M.V.K., J.S.S., F.A.H., Writing - original draft: R.R., T.K., T.E., C.A.G., D.G., M.K.I., M.V.K., P.L., S.W., F.A.H.; Writing – review \& editing: R.R., T.K., T.E., C.A.G., D.G., M.K.I., M.V.K., P.L., S.W., A.D., F.A.H.
\paragraph*{Competing interests:}
There are no competing interests to declare.

\newpage

\newpage

\appendix

\numberwithin{equation}{section} 
\numberwithin{figure}{section} 
\numberwithin{table}{section} 
\renewcommand{\thesection}{S}
\renewcommand{\thesubsection}{S.\arabic{subsection}}
\renewcommand{\thesubsubsection}{S.\arabic{subsection}.\arabic{subsubsection}}
\renewcommand{\thepage}{S\arabic{page}}
\setcounter{figure}{0}
\setcounter{table}{0}
\setcounter{equation}{0}
\setcounter{page}{1} %
\begin{center}
\section*{Supplementary Materials for\\ \scititle}
\end{center}
\tableofcontents
\newpage

\subsection{Data generation details} %
\label{sec:data_gen_appendix}

\subsubsection{Datasets}
The QM9 dataset~\cite{ruddigkeit2012enumeration, ramakrishnan2014quantum} is a collection of 133885 molecules with relaxed geometries and stoichiometry C\textsubscript{$c$}H\textsubscript{$h$}N\textsubscript{$n$}O\textsubscript{$o$}F\textsubscript{$f$} with $c,h,n,o,f \geq 0$ and $c+n+o+f \leq 9$.
We split the dataset randomly in an 80:10:10 ratio for training, validation and testing.

The extrapolation capabilities of STRUCTURES25 are tested on the QMugs dataset~\cite{isert2022qmugs}. The latter contains more than 665k drug-like molecules from the ChEMBL database. We filter out sulfur, chlorine, bromine and iodine since these elements do not appear in the QM9 dataset. For comparison with M-OFDFT~\cite{zhang2024overcoming}, we split the dataset into multiple bins according to size. The first bin comprises molecules with 10 to 15 heavy atoms and is used for training. The following bins have a width of 5 heavy atoms and contain 50 randomly sampled molecules each which are used in density optimization.

\subsubsection{Kohn-Sham DFT settings}\label{sec:kohn_sham_setting}
For the training and test label generation, KS-DFT calculations were carried out using the open source software package \textsc{PySCF}~\cite{sun2020recent, sun2018pyscf, Sun2015}. For ease of comparison with prior work M-OFDFT~\cite{zhang2024overcoming}, we largely choose identical hyperparameters. Restricted-spin calculations employing the \mbox{6-31G(2df,p)} basis set \cite{krishnan1980a} were conducted using the established general gradient approximation (GGA) functional PBE~\cite{perdew1996generalized} with a grid level of~3 for QM9 and a grid level of~2 for QMugs, respectively.
For larger molecules with more than 30 atoms, density fitting with the \mbox{def2-universal-jfit} basis set was enabled.
We set the convergence tolerance to the \textsc{PySCF} default of $2.72 \cdot 10^{-5}$~meV for QM9 and to 1~meV for QMugs. We use the commutator direct inversion of the iterative subspace (C-DIIS) ~\cite{pulay1982improved, kudin2002black} method with a maximal subspace of eight iterations. Minimal atomic orbitals (MINAO) was used as initialization~\cite{almlf1982principle, vanlenthe2006superposition}. The open source nature and python implementation of \textsc{PySCF} enabled us to insert callbacks in between the SCF steps of the KS-DFT computation. These serve two purposes: First, to extract density labels, energy labels, gradient labels and DIIS coefficients. Secondly and implicitly, to add perturbations to the Fock matrix, slowing down the convergence and leading to more varied training labels (see section~\ref{sec:pertubations_appendix}).

In KS-DFT, so-called Kohn-Sham orbitals $\phi_i:\mathbb{R}^3\rightarrow\mathbb{R}, i\in 1,\dots,N_\mathrm{KS}$ are introduced.  These orbitals describe non-interacting electrons in an effective potential $V_\mathrm{eff}$ such that the resulting ground state energy and density exactly match the interacting system. This approach leads to the well known Kohn-Sham equations~\cite{kohn1965self}
\begin{equation} \label{eq:kohn_sham_equations}
    \left(-\frac{1}{2}\nabla^2 + V_\mathrm{eff}\left[ \{\phi_i\} \right]\right) \phi_i = \varepsilon_i \phi_i
\end{equation}
which need to be solved iteratively, refining the orbitals and the effective potential they generate until self-consistency is achieved.

In molecules, the Kohn-Sham orbitals are expressed as a linear combination of atomic basis functions $\{\eta_\alpha\}_{\alpha\in 1,\dots,N_\mathrm{KS}}$ according to $\phi_i(\mathbf r) = \sum_{\alpha} C_{\alpha i} \eta_\alpha(\mathbf r)$.
In this representation, the Kohn-Sham equations at iteration $t$ are given by
\begin{equation}\label{eq:kohn_sham_in_basis}
    \mathbf{F}^t \mathbf{C}^t = \mathbf{S} \mathbf{C}^t \boldsymbol{\varepsilon}^t,
    \qquad F_{\alpha \beta}^t = \int \mathrm{d}^3 \mathbf{r}  \, \eta_\alpha(\mathbf r) \left[-\frac{1}{2}\nabla^2 + V_\mathrm{eff}[\{\phi^{t-1}_i\}](\mathbf r)\right] \eta_\beta(\mathbf r),
\end{equation}
where $S_{\alpha \beta} = \int \mathrm{d}^3 r \, \eta_\alpha(\mathbf r) \eta_\beta(\mathbf r)$ is the overlap matrix and $\boldsymbol{\varepsilon}^t$ the diagonal matrix of eigenvalues $\varepsilon^t_1$ to $\varepsilon^t_{N_\mathrm{KS}}$.
A naive implementation of two-electron integrals has a time complexity of $\mathcal O(N^4)$ with system size, which can be reduced to cubic scaling by employing density fitting~\cite{whitten1973coulombic, dunlap1979some, vahtras1993integral}. Solving the generalized eigenvalue problem also comes with a time complexity of $\mathcal O(N^3)$. This scaling impedes the application of Kohn-Sham DFT to larger systems.

In our linear orbital-free ansatz, we use a different basis set $\{\omega_\mu\}_{\mu\in 1,\dots,N_\mathrm{OF}}$ to directly represent the density as a linear combination 
\begin{equation}
    \rho_\mathrm{OF}(\mathbf r) = \sum_{\mu= 1}^{N_\mathrm{OF}} p_{\mu} \omega_\mu(\mathbf r),
\end{equation}
where $p_\mu$ are new coefficients that can be obtained from $C_{\alpha i}$ using density fitting as described in section~\ref{sec:label_generation}.

\subsubsection{Perturbation of the effective potential}\label{sec:pertubations_appendix}
Our principal aim was to overcome a fundamental limitation of prior work~\cite{zhang2024overcoming, ryczko2022orbitalfree} and achieve convergent density optimization, an essential quality for application of the second Hohenberg-Kohn theorem. This work shows that a well-behaved functional can be obtained when training on more varied data. Previous work~\cite{zhang2024overcoming} has shown how to generate training data using Kohn-Sham DFT, where for each SCF iteration one electron density, together with its energy and gradient labels were extracted. This approach has a drawback: The resulting labels are poorly distributed around the ground state, to the point where there are gaps in the difference between sample and ground state kinetic energy where almost no samples are generated (see Fig.~3C in the main text). More successful training of a machine learning model depends on labeled electron densities well distributed around the ground state density. It is however not straightforward to generate energy and gradient labels for a given density directly. This would require using an inverse Kohn-Sham approach, which unfortunately is still a ``numerical minefield''~\cite{shi2021inverse}.

Our main contribution is to instead perturb the effective potential \(V_{\mathrm{eff}}\), which is used in each SCF iteration to generate the electron density of the next iteration (cf.~ Eqs.~\ref{eq:kohn_sham_in_basis} and~\ref{eq:kohn_sham_in_basis_perturbed}). This approach is simple and stable, and results in labeled data that is much more evenly distributed than densities generated naively from Kohn-Sham DFT. It also offers direct control of the strength of perturbation in the effective potential \(V_{\mathrm{eff}}\) which in turn correlates with the strength of perturbation in electron densities and energies. 
This is illustrated in Fig.~3C, where the difference between the non-interacting kinetic energy \(T_\mathrm{S}\) and the corresponding value at the ground state \(T_\mathrm{S}^{\mathrm{gs}}\) is shown, using both perturbed and unperturbed \(V_{\mathrm{eff}}\). In both cases, samples are generated from all molecular geometries in the QM9 validation set, the higher total number of samples in the former case stems from the increased number of SCF iterations per molecule due to the perturbations.

The effective potential \(V_\mathrm{eff}\) in Eq.~\ref{eq:kohn_sham_equations} is perturbed from the sixth up to the 26th SCF iteration, counting the initial guess as the zeroth iteration. Only these perturbed samples are used in training.
The perturbed Kohn-Sham equations read \(\left[-\frac{1}{2}\nabla^2 + V_\mathrm{eff}[\{\phi_i^{t-1}\}] + \Delta^t \right] \phi^t_i = \varepsilon^t_i \phi^t_i\), with the perturbation function \(\Delta(\mathbf r) = \sum_\mu d_\mu \omega_\mu(\mathbf r)\), its coefficients \(d_\mu\) are sampled from a normal distribution with a standard deviation decreasing linearly from \(0.102\) to \(0.002\) over the SCF iterations and then multiplied with the orbital-free basis functions $\{\omega_\mu\}$.
In the \(\{\eta_\alpha\}\) basis representation this amounts to 
\begin{align}\label{eq:kohn_sham_in_basis_perturbed}
    \left( \mathbf{F}^t + \mathbf{\Delta}^t \right) \mathbf{C}^t &= \mathbf{S} \mathbf{C}^t \boldsymbol{\varepsilon}^t \\
    \Delta^t_{\alpha \beta} &= \sum_\mu d^t_\mu \int \mathrm{d}^3 r  \, \eta_\alpha(\mathbf r) \eta_\beta(\mathbf r) \omega_\mu(\mathbf r).
\end{align}
Given that all Kohn-Sham DFT calculations are performed in this matrix representation one might wonder why we do not sample \(\mathbf{\Delta}\) directly. This however could lead to inconsistent perturbed Fock matrices \(\mathbf{F}^t + \mathbf{\Delta}^t\) that cannot be generated by some operator of the form \(-\frac{1}{2}\nabla^2 + V_\mathrm{eff}\left[\{ \phi^{t-1}_i \}_{i\in 1,\dots,N_\mathrm{KS}}\right]\).

\subsubsection{Label generation}
\label{sec:label_generation}
To efficiently learn the energy functional, we create tuples of the molecular structure $\mathcal M$, the density coefficients $\mathbf{p}$, the target energy $E_\mathrm{target}$ and the gradient of the target with respect to the coefficients $\nabla_\mathbf{p} E_\mathrm{target}$.

To obtain density coefficients in the orbital-free ansatz (Eq.~2 in the main text), we employ density fitting.
The \textit{resolution of identity} method by Whitten\cite{whitten1973coulombic} and Dunlap\cite{dunlap1979some} can be used to fit the orbital-free density coefficients $\mathbf{p} = \{p_\mu\}_{\mu\in 1,\dots,N_\mathrm{OF}}$ to a Kohn Sham density. Here, the discrepancy of the fit is measured by the residual Hartree energy
\begin{align}
    E_\mathrm{H} [\rho_\mathrm{KS}-\rho_\mathrm{OF}] 
    &= \int \mathrm d \mathbf{r}\int \mathrm d \mathbf{r'}\frac{(\rho_\mathrm{KS}(\mathbf r)-\rho_\mathrm{OF}(\mathbf r))(\rho_\mathrm{KS}(\mathbf{ r'})-\rho_\mathrm{OF}(\mathbf{ r'}))}{|\mathbf r -\mathbf{ r'}|} \\
    &= \mathbf{p}^\top \mathbf{\tilde W} \mathbf{p} - 2 \mathbf{p} \,\mathrm{tr} (\mathbf{\tilde L} \mathbf\Gamma)+ \sum\limits_{\alpha\beta\gamma\delta} \Gamma_{\alpha\beta} \tilde D_{\alpha\beta,\gamma\delta} \Gamma_{\gamma\delta}. \label{eq:hartree_diff}
\end{align}
Here $\tilde{W}_{\mu\nu} = (\omega_\mu |\omega_\nu)$, $\tilde{L}_{\mu,\alpha\beta} = (\omega_\mu |\eta_\alpha\eta_\beta)$ and $\tilde{D}_{\alpha\beta,\gamma\delta}  = (\eta_\alpha\eta_\beta|\eta_\gamma\eta_\delta)$ are the overlap matrices between the basis functions under the kernel $\frac{1}{|\mathbf r-\mathbf{ r'}|}$, where we define
\begin{equation}
    (f|g)= \int \mathrm d \mathbf r \int \mathrm d \mathbf r' \frac{f(\mathbf r)g(\mathbf{ r'})}{|\mathbf r -\mathbf{ r'}|}.
\end{equation}
The trace $\mathrm{tr}$ sums over the indices of the Kohn Sham basis as in $[\mathrm{tr}(\mathbf{\tilde L} \mathbf\Gamma)]_\mu =\sum\limits_{\alpha\beta}\tilde L_{\mu,\alpha\beta} \Gamma_{\alpha\beta}$.
The density matrix $\Gamma_{\alpha\beta} = \sum\limits_i C_{i\alpha} \, n_i \, C_{i\beta}$ is obtained by contracting the orbitals with the corresponding occupation number $n_i$.

As Eq.~\ref{eq:hartree_diff} is a quadratic form in $\mathbf{p}$, it can be minimized analytically. But in agreement with M-OFDFT\cite{zhang2024overcoming}, we found that also considering the residual external energy in the minimization yields better fitted external and exchange correlation energies, as using the residual Hartree energy alone only leads to a close fit for this energy. The residual external energy reads
\begin{align}
    E_\mathrm{ext}[\rho_\mathrm{KS}-\rho_\mathrm{OF}] &= \int \mathrm d \mathbf{r}\sum\limits_{(Z_a, \mathbf{R}_a) \in \mathcal{M}} \frac{-Z_a(\rho_\mathrm{KS}(\mathbf{r})-\rho_\mathrm{OF}(\mathbf{r}))}{|\mathbf{R}_a-\mathbf{r}|}
    \\ &= \mathbf v_\mathrm{ext}^\top \,\mathbf{p}-\mathrm{tr}(\Gamma \mathbf{V}_\mathrm{ext}),\label{eq:external_diff}
\end{align}
with $[{\mathbf v_\mathrm{ext}}]_\mu = E_\mathrm{ext}[\omega_\mu]$ and $[{\mathbf V_\mathrm{ext}}]_{\alpha\beta} = E_\mathrm{ext}[\eta_\alpha\cdot\eta_\beta]$ being the external energies of the individual basis functions and $\mathcal{M}$ denoting the molecule geometry, comprising nuclear positions $\mathbf{R}_a$ and corresponding charges $Z_a$.
The objectives of $E_\mathrm{ext}[\rho_\mathrm{KS}-\rho_\mathrm{OF}]=0$ and minimizing $E_\mathrm{H} [\rho_\mathrm{KS}-\rho_\mathrm{OF}]$ can be combined by differentiating Eq.~\ref{eq:hartree_diff} with respect to $\mathbf{p}$ and coupling with Eq.~\ref{eq:external_diff}, yielding an overdetermined linear system\footnote{
    The M-OFDFT supplementary\cite{zhang2024overcoming} suggests minimizing the loss
    \begin{equation}
        \mathcal{L}(\mathbf{p}) = E_H[\rho_\mathrm{OF}(\mathbf{p})-\rho_\mathrm{KS}] + \left(E_\mathrm{ext}[\rho_\mathrm{OF}(\mathbf{p})]-E_\mathrm{ext}[\rho_\mathrm{KS}]\right)^2,    
    \end{equation}
    but the available code optimizes the least squares problem as we do.
} which is solved by minimizing
\begin{equation}
    \mathcal{L}(\mathbf{p}) =\left\lVert
    \begin{pmatrix}
    \mathbf{\tilde W} \\
    \mathbf v_\mathrm{ext}^\top
    \end{pmatrix}
    \mathbf{p}
    -
    \begin{pmatrix}
    \mathrm{tr}(\mathbf{\tilde{L}} \mathbf\Gamma) \\
    \mathrm{tr}(\mathbf\Gamma \mathbf{V}_\mathrm{ext})
    \end{pmatrix}
    \right\lVert^2
\end{equation}
using a least squares solver.

To obtain labels for the energy targets, we write the total energy as the sum
\begin{equation}
    E_\mathrm{tot} = T_\mathrm{S} + E_\mathrm{XC} + E_\mathrm{H} + E_\mathrm{ext} + E_\mathrm{nuc},
\end{equation}
where the nuclear repulsion energy $E_\mathrm{nuc}$ only depends on the molecular structure $\mathcal M$. The Hartree energy $E_\mathrm H$ as well as the external energy $E_\mathrm{ext}$ are known functionals of the density. There are many popular approximations of the exchange-correlation functional $E_\mathrm{XC}$ that are \emph{pure}, meaning that they can also be computed from just the density.
The non-interacting kinetic energy is only available in the Kohn-Sham setting via
\begin{align}
    T_\mathrm{S}(\mathbf{C}) = \sum\limits_i\langle \phi_i | \hat T | \phi_i \rangle = \sum_{\alpha, \beta} \sum_i C_{\alpha i} C_{\beta i} \langle \eta_\alpha | \hat T | \eta_\beta\rangle.
\end{align}
Following prior work~\cite{zhang2024overcoming}, we calculate the kinetic energy label for the orbital-free density such that the total energy remains constant, $E_\mathrm{tot}(\mathbf p) = E_\mathrm{tot}(\mathbf C)$, yielding
\begin{equation}
    T_\mathrm{S}(\mathbf{p}) = T_\mathrm{S}(\mathbf{C}) + E_\mathrm{eff}(\mathbf{C})  - E_\mathrm{eff}(\mathbf{p}), \qquad E_\mathrm{eff} \coloneq E_\mathrm{H} + E_\mathrm{XC} + E_\mathrm{ext},
\end{equation}
which helps mitigate errors introduced during density fitting.

Finally, we need the gradients of the energy contributions as a training signal for the machine learning model. All contributions except for the kinetic energy can be simply differentiated with respect to the density coefficients $\mathbf p$ to directly obtain the corresponding label.
The gradient of the non-interacting kinetic energy can be calculated by using the fact that the resulting density of each individual SCF iteration is the ground state of the non-interacting system given by the Fock operator
\begin{equation}
    \hat F^t = \hat T_\mathrm{S} + \hat V_\mathrm{eff}^t,
\end{equation}
where $\hat V_\mathrm{eff}^t = \hat V_\mathrm{eff}[\rho^{t-1}]$ is the effective potential generated by the previous density $\rho^{t-1}$.
At the ground state of the non-interacting system, the functional derivative of the total energy with respect to the electron density, subject to the conservation of electron number, vanishes. The optimality condition $\frac{\delta E}{\delta \rho(\mathbf r)} = \mu$ leads to the identity
\begin{equation}
    \frac{\delta T_\mathrm{S} [\rho]}{\delta \rho(\mathbf r)} = - V_\mathrm{eff}(\mathbf r) - \Delta(\mathbf r) + \mu,
\end{equation}
where the constant $\mu$ is the chemical potential, and where we include the perturbation function $\Delta(\mathbf r)$ from above. Upon integrating over the density basis functions, we obtain
\begin{equation}
    \nabla_{p_\nu} T_\mathrm{S}(\mathbf{p})
    = \int \frac{\delta T_\mathrm{S}[\rho]}{\delta \rho(\mathbf{r})} \omega_\nu(\mathbf{r}) \mathrm d \mathbf r
    = \int \left( -V_\mathrm{eff}(\mathbf r)  - \Delta(\mathbf r) + \mu \right) \omega_\nu(\mathbf{r}) \mathrm d\mathbf{r}.
\end{equation}
The unknown chemical potential $\mu$ can be set to zero, as this only yields a gradient contribution orthogonal to the manifold of normalized densities, which is projected out in density optimization (see section~\ref{sec:density_optimization}).  Instead of obtaining the effective potential function $V_\mathrm{eff}(\mathbf r)$ from the coefficients in the orbital basis, we follow M-OFDFT~\cite{zhang2024overcoming} and directly use the effective potential vector given by
\begin{equation}
    \mathbf{v}_\mathrm{eff}(\mathbf{p}) = \nabla_\mathbf{p} \left(E_\mathrm{H}(\mathbf{p}) + E_\mathrm{XC}(\mathbf{p}) + E_\mathrm{ext}(\mathbf{p}) \right).
\end{equation}
Substituting $\Delta(\mathbf r) = \sum_\mu d_\mu \omega_\mu(\mathbf r)$, we obtain the gradient of the kinetic energy via
\begin{equation}
    \nabla_\mathbf{p} T_\mathrm{S}(\mathbf{p}) = -{\mathbf{v}}_\mathrm{eff} - \mathbf{W} \mathbf{d},
\end{equation}
where $W_{\mu \nu} = \int \mathrm d \mathbf r\,\omega_\mu (\mathbf{r}) \omega_\nu (\mathbf{r})$ is the overlap matrix of density basis functions.

In practice, the direct inversion of the iterative subspace (DIIS) is used to accelerate the convergence of the SCF iterations. Using DIIS, the new Fock matrix is a weighted sum of the previous Fock matrices,
\begin{equation}
    \mathbf{\tilde F}^t = \sum_{\tau=1}^t \pi_\tau^t \mathbf F^\tau, \qquad \sum_{\tau = 1}^t \pi_\tau^t = 1.
\end{equation}
Thus, the effective potential in iteration $t$ is given by
\begin{equation}
    \mathbf{\tilde V}_\mathrm{eff}^t = \sum_{\tau = 1}^t \pi_\tau^t \mathbf V^\tau_\mathrm{eff},
\end{equation}
and we need to replace the effective potential vector above with
\begin{equation}
    \tilde{\mathbf v}^t_\mathrm{eff} = \sum_{\tau = 1}^t \pi_\tau^t \mathbf v^\tau_\mathrm{eff}.
\end{equation}
Taken together, the above equations describe how to obtain density, energy and gradient labels from perturbed KS-DFT SCF iterations.

\subsection{Model overview} %
\label{subsec:model_overview}

The model input consists of a molecular graph with atomic positions $\{\mathbf R_a\}$ and atom types $\{Z_a\}$ as well as the coefficients $\mathbf p$, representing the electron density in terms of atom-centered basis functions, cf.~Eq.~2 in the main text. As is customary in local canonicalization, we compute an equivariant local coordinate system for each atom based on the relative position of adjacent non-hydrogen atoms. 
The basis functions $\{\omega_\mu\}$ are a product of a radial function and spherical harmonics. As a consequence, the density coefficients transform via Wigner-D matrices under 3D rotations. To achieve invariance w.r.t.~the global orientation of the molecule, the coefficients are transformed into the respective local frame at each atom. As in M-OFDFT\cite{zhang2024overcoming}, the coefficients undergo a ``natural reparametrization'' into an orthonormal basis, that is specific to the molecule geometry (see section~\ref{sec:enhancement_modules} for details); and the coefficients are rescaled dimensionwise to standardize the model input and the desired gradient range of the model w.r.t.~the input coefficients.

The preprocessed atom-wise coefficients are embedded as node features with a feature dimension of 768.  To this end, they are first passed through a shrink gate module
\begin{equation}
    \text{ShrinkGate}(\tilde{\mathbf{p}}) = \lambda_\mathrm{out} \tanh(\lambda_\mathrm{in} \,\tilde{\mathbf{p}}),
\end{equation}
with learnable parameters $\lambda_\mathrm{in}, \lambda_\mathrm{out}$ and subsequently through an MLP. 
Pairwise distances between nodes are embedded as edge features via a Gaussian basis function (GBF) module, given by
\begin{align}
    &\tilde{e}_{ij} = \eta_{\mathrm{mul}}(Z_i, Z_j)\|\mathbf{r}_i - \mathbf{r}_j\| + \eta_{\mathrm{bias}}(Z_i, Z_j)\,,\\
    &e_{ij}^k = \frac{1}{\sqrt{2\pi}\sigma_k}\exp\left(\frac{(\tilde{e}_{ij} - \mu_k)^2}{2\sigma_k^2}\right).
\end{align}
Here, $\eta_{\mathrm{mul}},\eta_{\mathrm{bias}}$ are learnable scalars, which depend on the atom types of sending and receiving node, $k \in \{0,1,\dots, 127\}$ and $\mu_k, \sigma_k$ are learnable mean and standard deviation of the $k$-th Gaussian, respectively. We initialize $\eta_{\mathrm{mul}}$ to 1 and $\eta_{\mathrm{bias}}$ to 0, while $\mu_k$ and $\sigma_k$ are drawn from a uniform distribution in the interval $[0,3]$.

Further, an embedding of the atom number $Z$ and an aggregation of edge features over neighboring nodes $\operatorname{MLP}\left(\sum_{j \in \mathcal{N}(i)}e_{ij}^k\right)$ are added to the node features. These are then passed to a message-passing graph neural network. At the core of the architecture, we apply 4 (or 8 for QMugs) Graphormer blocks~\cite{ying2021transformers}.
For experiments with local architectures, we propagate messages along a graph with a radial cutoff of 6 Bohr
and otherwise use the fully connected graph. Before each attention block, we also apply a node-wise layer norm.
Finally, we employ an energy MLP which produces an atom-specific energy contribution from the final node features. Combined with an atom-specific contribution (``atomic reference module'', see below) based on the statistics of the data, the individual energies of each molecule are aggregated by summation into the total energy prediction.
For the QMugs model, we used a hierarchical energy readout (similar to prior work~\cite{batatia2022mace}), where, instead of a single energy MLP after the final layer, we employ an energy MLP after every second transformer block to predict atom-specific energy contributions. In the end, all energy contributions of the individual readouts are summed into the final atom-wise energies. 
Having readout MLPs also in intermediate layers, where the field of view is still small, enforces the prediction of more local energy contributions whereas MLPs after later layers are expected to capture increasingly non-local effects.

\paragraph{Tensorial messages} As stated in the main text, we use local frames not solely to canonicalize input density coefficients with respect to global rotations, but also adopt a recently introduced approach~\cite{lippmann2025beyond} to modify the Graphormer architecture. In this modification, the known relative orientation of local frames is additionally leveraged during message passing by transforming node features from one local frame to another, allowing for the exchange of ``tensorial'' messages. These enable the network to effectively communicate non-scalar geometric features -- something that would not be possible without the frame-to-frame transition.
The representations under which internal features, i.e., the queries, keys and values in the attention mechanism, transform are hyperparameters. We choose these features to consist of 513 scalars, which remain invariant under rotations, and 85 vectors.
We modify the attention mechanism of the Graphormer in the following way: when updating the features $f_i$ of a given node $i$, keys $k_j$ and values $v_j$ of any adjacent node $j$ are transformed into the local frame of node $i$ before computing the attention weights and aggregating the values:
\begin{align}\label{eq:tensorframes-attention}
    f_i^{(t+1)} &= \bigoplus_{j \in \mathcal{N}(i)} a\Big(q_i, \ R(g_i) R(g_j^{-1}) k_j,  \ \mathbf r_i - \mathbf r_j  \Big)  \ R(g_i) R(g_j^{-1}) v_j , \\
    \mathrm{ with } \quad  a(q, k, \mathbf{r}) &= \mathrm{soft max} \left( \frac{q \cdot k}{\sqrt{d}} + \mathrm{MLP}(\mathrm{GBF}(\| \mathbf r \|)) \right),
\end{align}
where $\mathcal{N}(i)$ is the neighborhood of node $i$. $R$ is the group representation of the keys and queries w.r.t.~$\mathrm{SO}(3)$-transformations (chosen to be the same). $g_i$ denotes the rotation from the global frame into the local frame of node $i$.
Our experiments indicate that incorporating tensorial messages, which enable direct communication of geometrical features, improves model performance (cf.~Table~\ref{tab:tensorial_messages}). 

\begin{table}[ht]
    \centering
    \caption{\textbf{Scalar vs. tensorial messages on QM9.}}
    \vspace{10pt}
    \begin{tabular}{cccccc}
        \thickhline
        Tensorial Messages & $|\Delta E|$ & $|\Delta E|/ N_A$ & $\| \Delta \rho \|_2$ & $\| \Delta \rho \|_{2} / N_e$ \\
        & (mHa) & (mHa) &  & ($10^{-4}$)\\
        \hline
        $\times$ & 0.73 & 0.044 & 0.022 & 3.3 \\
        \checkmark & 0.64 & 0.038 & 0.014 & 2.1 \\
        \thickhline
    \end{tabular}
    \label{tab:tensorial_messages}
\end{table}

\paragraph{Radial cutoff} 

For our experiments on QMugs we construct the molecular graph using a radial cutoff instead of working with the fully connected graph. More specifically, we remove all edges between atoms with a pairwise Euclidean distance of $d > d_{c}$. We choose $d_{c} = 6 \, \mathrm{Bohr}$ as default value. 
The primary reason for introducing a cutoff is that message passing on the fully connected graph scales as $\mathcal{O}(N^2)$ whereas a radial cutoff reduces the complexity to $\mathcal{O}(N)$. Additionally, we observe in our ablation study that training with cutoff leads to a better generalization of the model to larger molecules, see Table~1 in the main text.
Further model hyperparameters are listed in Table~\ref{tab:hparams}.

\begin{table}[ht]
    \centering
    \caption{\textbf{Default model hyperparameters for our modified version of Graphormer.}}
    \vspace{10pt}
    \begin{tabular}{@{\hskip 10pt} l c @{\hskip 10pt}}
        \thickhline
        \textbf{Hyperparameter} & \textbf{Value} \\
        \hline
        Number of layers       & 8 \\
        Attention heads        & 32 \\
        Node dimension       & 768 \\
        Irreps for keys and values & 513 scalars,\\
         & 85 vectors\\
        GBF dimension  & 128\\
        $\lambda_\mathrm{out}$ (Shrink gate) & 10\\
        $\lambda_\mathrm{in}$ (Shrink gate) & 0.02\\
        \thickhline
    \end{tabular}
    \label{tab:hparams}
\end{table}

\subsection{Enhancement modules}%
\label{sec:enhancement_modules}
Proper normalization of model inputs and targets is critical for successful training of neural networks. In our pipeline, energy gradients with respect to density coefficients $\nabla_{\vp}E_{\mathrm{target}}$ are predicted via back-propagation through our ML functional. This entails that the scales of input density coefficients and gradient labels are linked, as multiplying the former by some factor amounts to rescaling the latter by the inverse factor.
Following M-OFDFT~\cite{zhang2024overcoming}, we thus employ a number of enhancement modules to constrain energies, gradients and coefficients to favorable ranges.

Natural reparametrization enables the meaningful measurement of density and gradient differences. A difference $\Delta \mathbf{p}$ in density coefficients in the LCAB ansatz (Eq. 2 in the main text) results in a residual density $\Delta \rho(\mathbf{r})= \sum_{\mu} \Delta p_{\mu} \omega_\mu(\mathbf{r})$ with an $L_2$ norm of 
\begin{equation}
   \lVert \Delta \rho \rVert^2_2 = \Delta \mathbf{p}^\top \mathbf{W} \Delta \mathbf{p},
\end{equation}
where $\mathbf{W}$ is the density basis overlap matrix. 
Transforming coefficients by $\mathbf{p} \mapsto \tilde{\mathbf{p}} \coloneq \mathbf{M}^\top\mathbf{p}$, where $\mathbf{M}$ is a matrix square root of the overlap matrix, i.e.~ $\mathbf{M} \mathbf{M}^\top = \mathbf{W}$, leads to an expansion for the density difference in terms of coefficient differences $\Delta \tilde{\mathbf{p}}$ only, 
\begin{equation}
   \lVert \Delta \rho \rVert^2_2 
   = \lVert \Delta \tilde{\rho} \rVert^2_2 
   = \Delta \tilde{\mathbf{p}}^\top \Delta \tilde{\mathbf{p}}.
\end{equation}
This implies that, after transformation, individual coefficient dimensions now equally and independently contribute to changes in the density. 
We can solve for $\mathbf{M}$ by considering the eigendecomposition $\mathbf{W} = \mathbf{Q}\boldsymbol{\Lambda}\mathbf{Q}^\top$ of the overlap matrix. %
The solution to $\mathbf{M}\mathbf{M}^\top = \mathbf{W}$ has a rotational degree of freedom as any matrix $\mathbf{M} = \mathbf{Q}\boldsymbol{\Lambda}^{\!1 \!/ 2} \mathbf{O}$, with an arbitrary orthogonal matrix $\mathbf{O}$, results in the overlap matrix upon squaring. We here shortly discuss the implications of choosing between two very distinct choices for $\mathbf{O}$.
Natural reparametrization is not only carried out on the coefficients $\mathbf{p}$ but also on all related quantities like gradients and basis functions. For the density $\rho(\mathbf{r})$ to remain invariant under reparametrization, basis functions transform via $\omega_\mu \mapsto \tilde{\omega}_\mu = \sum_\nu M^{-1}_{\mu \nu} \omega_\nu.$
As this transformation orthonormalizes the basis functions, $\int \tilde{\omega}_\mu(\mathbf{r})\tilde{\omega}_\nu(\mathbf{r}) = \delta_{\mu \nu}$, we stress the similarity to the non-orthogonality problem put forward by Löwdin~\cite{loewdin1950nonorthogonal, loewdin1970nonorthogonal}. This seminal work suggests choosing $\mathbf{O} = \mathbf{\mathbf{Q}^\top}$ and therefore $\mathbf{M}_{\mathrm{sym}} \coloneq \mathbf{Q} \boldsymbol{\Lambda}^{\! 1 \! / 2}\mathbf{Q}^\top$, the \textit{symmetric orthogonalization} -- often simply called \textit{Löwdin orthogonalization}~\cite{aiken1980onLoewdin, aiken1975Loewdinminimum}. This flavor of reparametrization results in the orthogonalized basis set that is least distant to the original basis set, as measured by the $L_2$-norm between basis functions. Furthermore, any rotation of the original basis set commutes with the orthogonalization; that is, the symmetrically reparametrized basis functions transform equivariantly under orthogonal transformations, which includes spatial rotations and permutation of basis functions. 
The latter property in particular motivates our choice to reparametrize coefficients by $\mathbf{M}_{\mathrm{sym}}$.

Zhang et al.~\cite{zhang2024overcoming} propose $\mathbf{O}=\mathbf{1}$, i.e., $\mathbf{M} = \mathbf{Q} \boldsymbol{\Lambda}^{\! 1 \! / 2}$ in their supplementary. This transformation, called \textit{canonical orthogonalization} by Löwdin~\cite{loewdin1970nonorthogonal}, results in highly delocalized orbitals. That is, each individual basis function aims to summarize the information in all untransformed orbitals subject to orthogonality. In our experiments, this reparametrization performed significantly worse than the symmetric version described above. 
This is because this type of reparametrization not only nullifies any previous transformation into local frames, but also breaks permutation invariance, rendering the energy prediction dependent on the order of atoms in the molecule. However, the code provided by M-OFDFT~\cite{zhang2024overcoming} employs the symmetric reparametrization as we do.

The dimension-wise rescaling and atomic reference modules are implemented as in prior work~\cite{zhang2024overcoming}.
Dimension-wise rescaling linearly transforms each density coefficient independently, trading off resulting coefficient and gradient scale for each component.
The atomic reference module adds a simple linear fit to the learned part of our functional. It thereby reduces the dynamic range of the predicted energy and effectively centers the gradient labels.

\subsection{Model training}\label{sec:model_training_appendix}

For both QM9 and QMugs, we train the model for 90 epochs with a batch size of 128. Over the course of all epochs, the learning rate is reduced from $7 \cdot 10^{-5}$ to $0$ using a cosine annealing schedule~\cite{loshchilov2016sgdr}. As optimizer, we use AdamW~\cite{loshchilov2018decoupled} with $\beta_1=0.95$, $\beta_2=0.99$ and a weight decay factor of $10^{-10}$. We do not use dropout. 
The training hyperparameters are summarized in Table~\ref{tab:training_hparams}.

We apply a loss to both the energy $E$ and its gradient $\nabla_{\mathbf{p}}E$. For the energies, a simple $L_1$ loss is used to compare the output to the labels. The gradient needs further attention, since it is only known in the hyperplane of normalized densities. To compare the derivative of the predicted energy with the gradient label $\mathbf g_\mathrm{label}$, we follow M-OFDFT~\cite{zhang2024overcoming} and apply an $L_1$ loss on the projected difference,
\begin{equation}
    \mathcal L_\mathrm{gradient} = \left\Vert \left(\mathbf I - \frac{\mathbf w \mathbf w^T}{\mathbf w^T \mathbf w}\right) \left( \mathbf \nabla_{\mathbf p} E - \mathbf g_\mathrm{label}\right) \right\Vert_1,
\end{equation}
where $\mathbf w$ is the vector of integrals over the density basis with components $w_\mu = \int \mathrm d^3 r\, \omega_\mu(\mathbf r)$ and the expression $\left(\mathbf I - \frac{\mathbf w \mathbf w^T}{\mathbf w^T \mathbf w}\right)$ corresponds to a projection onto the hyperplane orthogonal to $w$.

For direct comparison with prior work~\cite{zhang2024overcoming}, we group the QMugs dataset by the number of heavy atoms into bins of width 5. The first bin contains molecules with $10-15$ heavy atoms and molecules with fewer heavy atoms are discarded from the set.  When training on the QMugs dataset, we first combine the QM9 dataset with the first QMugs bin (10-15 heavy atoms) for the initial training set. Following this initial training on the combined dataset, we perform an additional fine-tuning step. This fine-tuning is conducted exclusively on the QMugs subset containing molecules with 10-15 heavy atoms (the first bin). The fine-tuning uses the same hyperparameters as the initial training, except for the learning rate and epoch count. We start with a learning rate of $1 \cdot 10^{-5}$, which is also reduced to $0$ following a cosine annealing schedule over 30 epochs. All other training parameters, including the optimizer, loss function, and batch size, remain unchanged. Training on the QM9 dataset alone does not involve this fine-tuning step.

\begin{table}[ht]
    \centering
    \caption{\textbf{Default training hyperparameters.}}
    \vspace{10pt}
    \begin{tabular}{@{\hskip 10pt} l c @{\hskip 10pt}}
        \thickhline
        \textbf{Hyperparameter} & \textbf{Value} \\
        \hline
        Number of epochs & 90 \\
        Batch size & 128 \\
        Learning rate & $7 \cdot 10^{-5}$\\
        Adam $\beta_1$ & $0.95$\\
        Adam $\beta_2$ & $0.99$\\
        Weight decay & $10^{-10}$\\
        \thickhline
    \end{tabular}
    \label{tab:training_hparams}
\end{table}

\subsection[Initial electron density guess: Data-driven sum of atomic densities~(dSAD)]{\hspace{-0.3cm} Initial electron density guess: Data-driven sum of atomic densities~(dSAD)}\label{sec:dSAD_appendix}
A multitude of established methods for generating initial guesses for the electron density in KS-DFT exist, such as the MINAO initialization~\cite{almlf1982principle,vanlenthe2006superposition}. 
However, while it is cheap compared to the Kohn-Sham iterations because of the minimal basis that it uses, it still scales cubically with system size, and additionally requires density-fitting to transform the guess to the density basis~(Eq.~2 in the main text).

A much simpler and linear scaling option is a superposition of atomic densities (SAD). 
We have datasets with ground state density coefficients at hand, as these are required for training. This allows us to determine average atomic densities with a data-driven approach:
We take all instances of each atom type (i.e.~chemical element) in the dataset, and take the average of the corresponding coefficients over all these instances. 
Averages for coefficients corresponding to basis functions with $l>0$ are set to zero.
For a given molecule $\mathcal{M}$, concatenating these averages for all atom types in the molecule then yields $\bar{\vp}$, our data-driven SAD (dSAD).
However, this superposition of atomic densities is not necessarily normalized to the correct electron number. 
Since the number of electrons stays invariant during density optimization (see section~\ref{sec:density_optimization}), we need to normalize to generate a valid guess.
One approach is to uniformly scale the coefficients linearly to the correct electron number $\Ne$, leading to
\begin{align}
    \bar\vp_{\mathrm{uniform}} = \bar{\vp} \, \frac{\Ne}{\vw^{\top} \bar{\vp}} \,.
\end{align}
with the basis integrals $\vw$, see section~\ref{sec:model_training_appendix}.
While simple, this has a major shortcoming: 
The largest part of the electron density lies close to the cores, and, for elements other than hydrogen, this core density varies only very little between different instances of the same atom type in neutral molecules. Thus, the SAD guess describes the core very precisely. 
The coefficients of the inner $l=0$ basis functions largely describe this core density and should hence be varied very little in the normalization. 
However, scaling all coefficients by the same factor to achieve normalization does not respect this. 
For example, the core density of atomic species with high electronegativity (whose corresponding coefficients, on average, describe a higher number of electrons than their atomic number indicates), would be scaled down and hence underestimated. 

\begin{figure}[ht]
    \centering
    \includegraphics[width=0.5\textwidth]{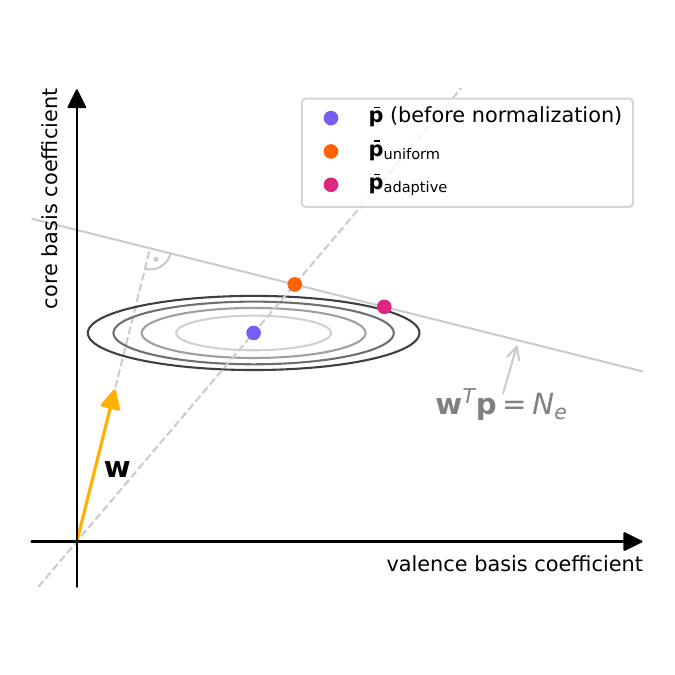}
    \caption{\textbf{Different methods for normalizing atomic densities.} 
    When extracting average atomic densities from the training set, these are not correctly normalized. Simple uniform scaling ($\bar\vp_{\mathrm{uniform}}$) neglects the relative invariance of core electrons under chemical bonding. The heteroscedastic estimate $\bar\vp_{\mathrm{adaptive}}$ takes the variability of different kinds of coefficients into account, see section~\ref{sec:dSAD_appendix}.}
    \label{fig:sad_normalization}
\end{figure}

This is why we propose a heteroscedastic normalization procedure, which adapts to the variance of the coefficients over the dataset: Coefficients with high variance are scaled more than those with low variance, as they are more likely to be far from the mean.
As the mean $\bar{\vp}$ and variance $\sigma_{\mu}$ of each coefficient are known, we can formulate this as an weighed least squares optimization problem with a linear constraint, where the weights of the squared deviations from $\bar{\vp}$ are given by the inverse squares of the variances 
(see also Fig.~\ref{fig:sad_normalization}):
\begin{align}
\bar\vp_{\mathrm{adaptive}} &= \argmax_{\vp,\,\, \vw^{\top}\vp = \Ne} \sum_{\mu} \frac{(p_{\mu} - \bar{p}_{\mu})^2}{2\sigma_{\mu}^2} = \bar{\vp} + \argmax_{\vd,\,\, \vw^{\top}\vd = \Delta \Ne} \sum_{\mu} \frac{d_{\mu}^2}{2\sigma_{\mu}^2} \label{eq:p_normalized_first}
\end{align}
with $\Delta \Ne = \Ne - \vw^{\top} \bar{\vp}$, the difference between the desired electron number and the one corresponding to the mean coefficients.
Introducing a Lagrange-multiplier $\lambda$, we get:
\begin{align}
    \mathcal{L}(\vd, \lambda) = \sum_{\mu} \frac{d_{\mu}^2}{2\sigma_{\mu}^2} + \lambda\bigg( \Big( \sum_{\mu} w_{\mu} d_{\mu} \Big) - \Delta \Ne \bigg) \,
\end{align}
and can solve for $d$ and $\lambda$
\begin{align}
    d_{\mu} &= - \lambda \sigma_{\mu}^2 w_{\mu} \,, \quad \lambda = - \frac{\Delta \Ne}{\sum_{\mu} \sigma_{\mu}^2 w_{\mu}^2} \,
\end{align}
to find
\begin{align}
(\bar\vp_{\mathrm{adaptive}})_{\mu} = \bar{p}_{\mu} + \Delta \Ne \frac{\sigma_{\mu}^2 w_{\mu}}{\sum_{\nu} \sigma_{\nu}^2 w_{\nu}^2} \,.
\end{align}
This result matches the intuition established above: The correction to each component of the average coefficients $\bar{\vp}$ is proportional both to the variance of it over the dataset, and the weight of its corresponding basis function. 
An illustration of this method compared to simply scaling the guess is shown in Fig.~\ref{fig:sad_normalization}. 

One could either apply this normalization per molecule, or per chemical element (in the latter case, $\Ne$ denotes the number of electrons corresponding to the atom type).
We choose the latter for simplicity, normalizing the electron number per element. 
The guess is computed before applying natural reparametrization.
Comparing our dSAD guess to the MINAO guess on 1000 QM9 molecules in Table~\ref{tab:initial_guess_errors}, we find that it produces guesses which are slightly closer to the ground state.

\begin{table}[ht]
    \centering
    \caption[Density errors of initial guesses]{\textbf{Density errors of initial guesses.} We compare the accuracy of the MINAO~\cite{almlf1982principle,vanlenthe2006superposition}, Hückel~\cite{lehtola2019initialguess} and the proposed dSAD guess (section~\ref{sec:dSAD_appendix}). Shown is the mean of the $L_2$ density error to the ground state across 1000 molecules from the QM9 dataset.}
    \label{tab:initial_guess_errors}
    \vspace{10pt}
    \begin{tabular}{lccc}
        \thickhline
        Initial guess & $\norm{\rho_{\mathrm{guess}} - \rho^{\mathrm{gs}}}_2 / \Ne \ (10^{-4})$ 
        & Computational complexity \\ \hline
        dSAD      & \phantom{1}52 %
        & $\mathcal{O}(N)$ \\
        MINAO    & \phantom{1}64 %
        & $\mathcal{O}(N^3)$ \\
        Hückel & 122 %
        & $\mathcal{O}(N^3)$ \\ \thickhline
    \end{tabular}
\end{table}

\subsection{Density optimization}
\label{sec:density_optimization}
After initialization of the electron density using our dSAD guess (section~\ref{sec:dSAD_appendix}), the learned energy functional enables its iterative optimization in order to find the ground state density and the corresponding energy.
To conserve the number of electrons, we must not diverge from the hyperplane of normalized densities $\{\mathbf p \colon \mathbf w^\top \mathbf p = N_e \}$. We thus project the step $\mathbf u$ of the optimizer onto the hyperplane such that the density coefficients are updated according to 
\begin{align}
    \mathbf p^{t+1} = \mathbf p^t + \left(\mathbf I - \frac{\mathbf w \mathbf w^T}{\mathbf w^T \mathbf w}\right) \mathbf u.
\end{align}
We use gradient descent with momentum as our optimizer, with a learning rate of $0.003$ and a momentum of $0.9$ for QM9. For the QMugs dataset, we reduce the learning rate to $0.0015$. These parameters were tuned such that the density optimization shows a fast and robust convergence. The number of iterations is limited to 5000.

Note that while KS-DFT typically requires significantly fewer steps to converge, it relies on much more complex SCF iterations. 
While OF-DFT requires more steps, each is computationally inexpensive. This results in a significant net gain in efficiency, particularly for larger systems, see section \ref{sec:runtime_comparison} for details.

The model is never trained on molecules from the test set, and density optimization is performed on the test set only.
As Fig.~2B in the main text illustrates, density optimization on the STRUCTURES25 functional converges to gradient norms of $10^{-13}$~Ha for smaller molecules. This level of convergence is more precise than our labels are, due to imperfect density fitting and a finite convergence tolerance in the Kohn Sham calculations of our ground truth, see section~\ref{sec:kohn_sham_setting}. We thus stop density optimization when the gradient norm for an entire molecule falls below $10^{-4}$~Ha.

Regarding the definition of chemical accuracy, an error of 1~kcal\,mol$^{-1}$ is a widely used threshold for acceptable energy errors. As there is no widely used definition for chemical accuracy of an electron density, we propose to compare the densities produced by different XC functionals using the same (def2-TZVP) basis set \cite{weigend2005a}. At the  local density approximation (LDA) rung we use BN05\cite{boronski2005}. Generalized gradient approximations (GGA) are represented by PBE and B97\cite{Becke1997}. As Meta-GGA, incorporating the kinetic energy density, we use R2SCAN\cite{furness2020accurate}. The highest accuracy levels are achieved by the hybrid-GGAs PBE0\cite{Adamo1999}, B3LYP\cite{Lee1988} or a meta-hybrid-GGA with PW86 exchange and B95 correlation~\cite{Perdew1986,Becke1996}. The ground state densities of 1000 randomly sampled molecules from QM9 were computed using our default Kohn Sham settings for each of the above functionals. We then measured the density difference between the XC functional PBE\cite{perdew1996generalized}, which was used for data generation, and this assortment of functionals, with results shown in Table~\ref{tab:xc_functionals}. 
The density difference between PBE and methods at least at the GGA level is in the range of $7.2\cdot 10^{-4}$ to $1.3\cdot 10^{-3}$ electrons in the L2 norm, while the density difference to the less accurate LDA BN05 is around $4.1\cdot 10^{-3}$ electrons. Given the above deviations from PBE densities, and given that the PBE functional has been used in more than 200~000 publications to date, we here define ``chemically accurate densities'' at the level of PBE as all those that come within $7.2\cdot 10^{-4}$ electrons in the L2 norm on QM9-sized molecules.

\begin{table}[ht]
    \centering
    \caption{\textbf{Density differences between PBE\cite{perdew1996generalized} and other XC functionals.} The $L_2$ norm of the ground state density difference is evaluated on 1000 molecules randomly sampled from QM9.}
    \vspace{10pt}
    \begin{tabular}{llc}
        \thickhline
        XC functional type& XC functional name & $\norm{\Delta {\rho} }_2 / \Ne \ (10^{-4})$ \\ \hline 
        LDA & BN05\cite{boronski2005} & 41\phantom{.0} \\
        GGA & B97\cite{Becke1997} & 13\phantom{.0} \\ 
        Meta-GGA & R2SCAN\cite{furness2020accurate} & \phantom{0}8.7 \\
        Hybrid-GGA & PBE0\cite{Adamo1999} & \phantom{0}7.2 \\
        Hybrid-GGA & B3LYP\cite{Lee1988} & \phantom{0}7.4 \\
        Hybrid-Meta-GGA & PW86 B95\cite{Perdew1986,Becke1996} & 11\phantom{.0} \\ \thickhline
    \end{tabular}
    \vspace{10pt}
    \label{tab:xc_functionals}
\end{table}

\begin{figure}
    \centering
    \includegraphics[width=0.7\linewidth]{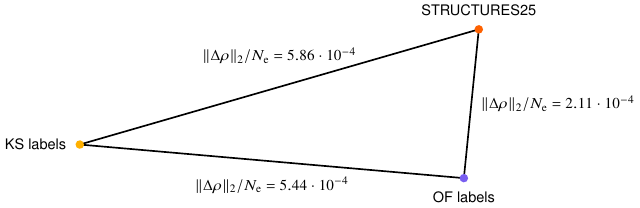}
    \caption{\textbf{The residual between STRUCTURES25 and its OF training labels is smaller than the error incurred by density fitting training labels to the KS calculations.} The sketch is to scale and illustrates the pairwise $L_2$ norm per electron between ground state densities, averaged over the QM9 test set.}
    \label{fig:norm-triangle}
\end{figure}

\subsection{QMugs trifluoromethoxy outliers} %

\begin{figure}[ht]
\centering
\includegraphics[width=0.6\textwidth]{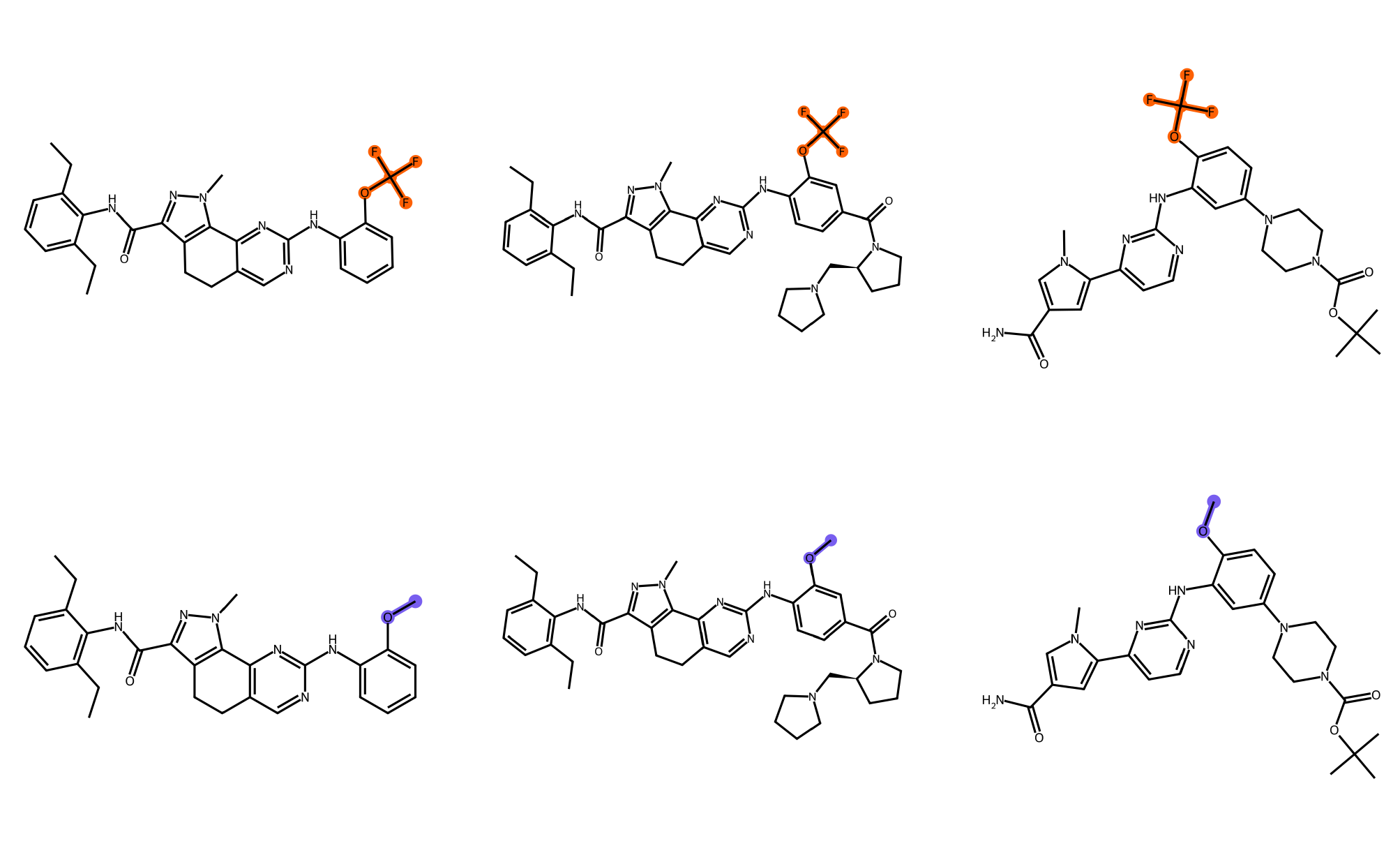}
\caption{\textbf{QMugs outliers with trifluoromethoxy group and their methoxy derivatives.} The three outliers in our QMugs test set with their trifluoromethoxy group marked orange and their methoxy derivatives marked blue. The latter have energy  errors in the usual range obtained for other QMugs samples. The experiment shows that the trifluoromethoxy groups, which are underrepresented in the training data, are the cause of the outliers.}
\label{fig:trifluoromethoxy}
\end{figure}

Fig.~4A in the main text, which shows the extrapolation accuracy from small to large organic molecules, reveals three outliers. It turned out that all three contained a trifluoromethoxy group, and that this chemical group was present only in the three conformers of a single molecule in the training set. We thus hypothesized that this group was responsible for the poor predictions.
To investigate this hypothesis, we substituted the fluorine atoms of the trifluoromethoxy group by hydrogen, yielding their methoxy derivatives (see Fig.~\ref{fig:trifluoromethoxy}). We re-optimized the geometry at the same level of theory as the original samples (GfN2-xtb using energy and gradient convergence criteria of $5\times 10^{-6}$~Ha and $10^{-3}$~Ha$\,\alpha^{-1}$)~\cite{isert2022qmugs}.  
Subsequent OF-DFT calculations employing the STRUCTURES25 functional (trained on QMugs) showed that the energy error per atom dropped from $11.69$~mHa to $0.4$~mHa%
, from $10.2$~mHa  to $0.2$~mHa%
, and  from $7.42$~mHa  to $0.47$~mHa %
respectively, well within the range of the other QMugs test samples. This indicates it was indeed the trifluoromethoxy group which caused the outliers. 

\begin{figure}
    \centering
    \includegraphics[width=0.7\linewidth]{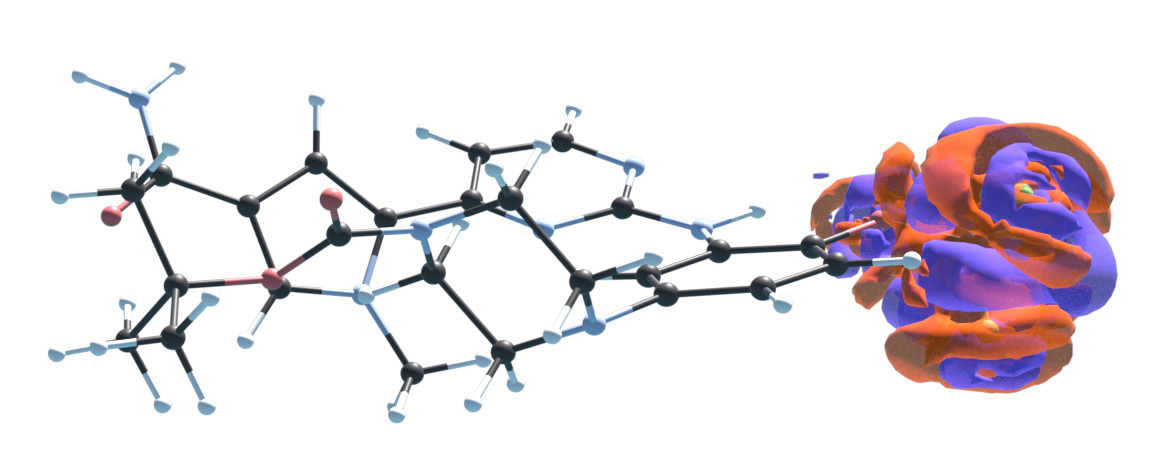}
    \caption{\textbf{For one of the energy prediction outliers from Fig.~4 in the main text: The density prediction error is fully localized around the trifluoromethoxy group. }  Isocontours of the density residual between STRUCTURES25 prediction and ground truth after density fitting.  The red/blue surfaces show positive/negative deviations above $0.02\,\mathrm{Bohr^{-3}}$ and the discrepancy is localized  on the trifluoromethoxy group on the right.}%
    \label{fig:trifluoro-density}
\end{figure}

To further analyze the phenomenon, we evaluated the accuracy of STRUCTURES25 on smaller molecules containing a trifluoromethoxy group. These exhibited large errors as well, see Table \ref{tab:trifluoro}. Finally, the density error for a larger molecule containing the group is visualized in Figure \ref{fig:trifluoro-density}, demonstrating that the error is spatially localized around the group in question. Taken together, these observations support the hypothesis that the lack of sufficiently many training molecules including the trifluoromethoxy group is the root cause of this failure mode.
Analyses of this type can help identify and close gaps in the training data.

\begin{table}[htb]
    \centering
    \caption{\textbf{Energy error of STRUCTURES25 relative to PBE/6-31G(2df,p) for small trifluoromethoxy molecules and their substitues.} For the substitues, the trifluoromethoxy moiety was replaced with a methoxy group. }
    \vspace{7pt}
    \begin{tabular}{lcc} \toprule
        Molecule & $\Delta E$ [mHa] & substitute $\Delta E$ [mHa] \\
        \midrule
        Trifluoro(methoxy)methane %
         & 629.21 & 1.57 \\
        (Trifluoromethoxy)benzene & 642.50 & 1.26 \\
        2-(Trifluoromethoxy)aniline & 639.28 & 1.81 \\
        \bottomrule
    \end{tabular}
    \label{tab:trifluoro}
\end{table}

\subsection{Negative densities}\label{sec:negative_densities} %

The representation of the density given in Eq.~2 in the main text in principle allows for regions of negative densities to occur. When using a trained model for density optimization this could lead to problems. Since negative densities are not physical and therefore no training data with negative densities exists, it is unclear if the model can make meaningful predictions for such non-physical densities.

\begin{figure}[ht]
    \centering
    \includegraphics{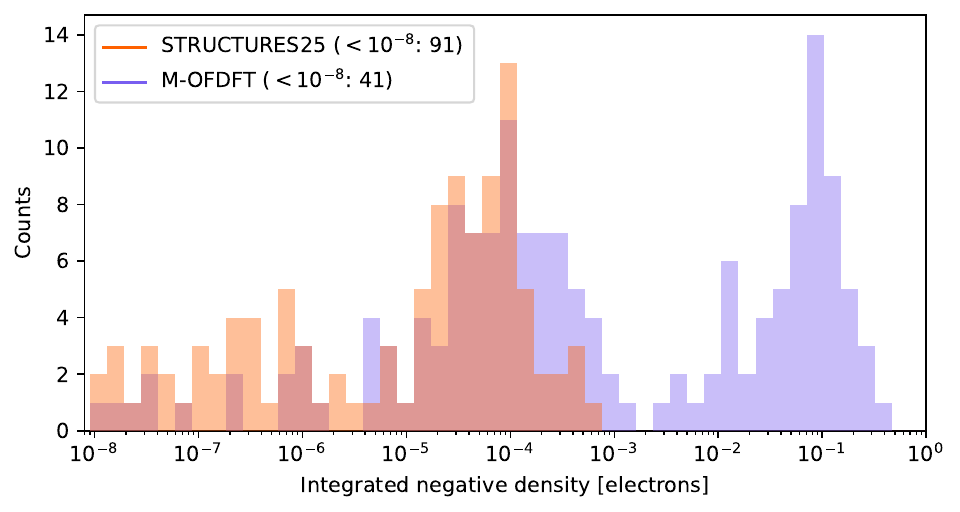}
    \caption{\textbf{Unphysical negative electron densities are admitted by the representation in Eq.~2 in the main text, but are not a problem in practice.} Shown is a histogram of integrated negative densities after density optimization for 200 random QM9 molecules and 6000 optimization steps. While negative density contributions are vanishingly small for STRUCTURES25, a number of densities optimized through the M-OFDFT functional contain significant negative contributions. The number of negative integrated densities below $10^{-8}$ electrons is not depicted but given in parentheses after the respective functional name.}
    \label{fig:negative_density_comparison} 
\end{figure}

In practice, we report that negative densities are no failure case for our models trained on the $E_{TXC}$ target. When starting from a reasonable initial guess such as dSAD (section~\ref{sec:dSAD_appendix}), the negative regions of the converged densities are insignificant. This is illustrated for our QM9 model and that of M-OFDFT~\cite{zhang2024overcoming} in Fig.~\ref{fig:negative_density_comparison}. If negative densities were to become a problem in the future, e.g. for larger molecules or when considering more elements, one could penalize negative densities directly in the density optimization. A penalization term of the form 
\begin{equation}
    L_{\mathrm{nd}} \coloneq \gamma \int \mathrm d \mathbf{r} \left( \max \left( - \rho(\mathbf{r}), 0 \right) \right)^2 ,
\end{equation}
can be added to the total energy, where \(\gamma > 0\) is a hyperparameter. Using this penalization term does not significantly change our results, which is why we do not use it for the reported experiments. Such a penalization term has the significant downside of requiring a grid for evaluation, which we otherwise do not need for the $E_\mathrm{TXC}$ target.

\subsection{Robustness with respect to initialization} %

Zhang et al.~\cite{zhang2024overcoming} require a precise machine-learned guess, ``ProjMINAO'', to produce their best results.
Their results deteriorate by an order of magnitude when initializing with Hückel~\cite{lehtola2019initialguess} densities and become worse still when initializing with a MINAO guess~\cite{zhang2024overcoming, almlf1982principle, vanlenthe2006superposition}.
To gauge the robustness of the STRUCTURES25 functional, we compare density optimizations initialized with our data-driven superposition of atomic densities (dSAD, section~\ref{sec:dSAD_appendix}) as well as MINAO and Hückel densities. Unlike prior work~\cite{zhang2024overcoming} we refrain from training a model to improve on the initially guessed density.

For the following comparisons, we randomly sample 1000 molecules from our QM9 test set. As a point of reference for the various initial guesses, we report their density error with respect to the ground state in table~\ref{tab:initial_guess_errors}. Predicted ground state density errors, $\rho^{\mathrm{gs}} - \rho^{\ast}_{\mathrm{guess}}$, from dSAD and MINAO are nearly identical while Hückel fares approximately an order of magnitude worse.

\begin{figure}[ht]
    \centering
    \includegraphics{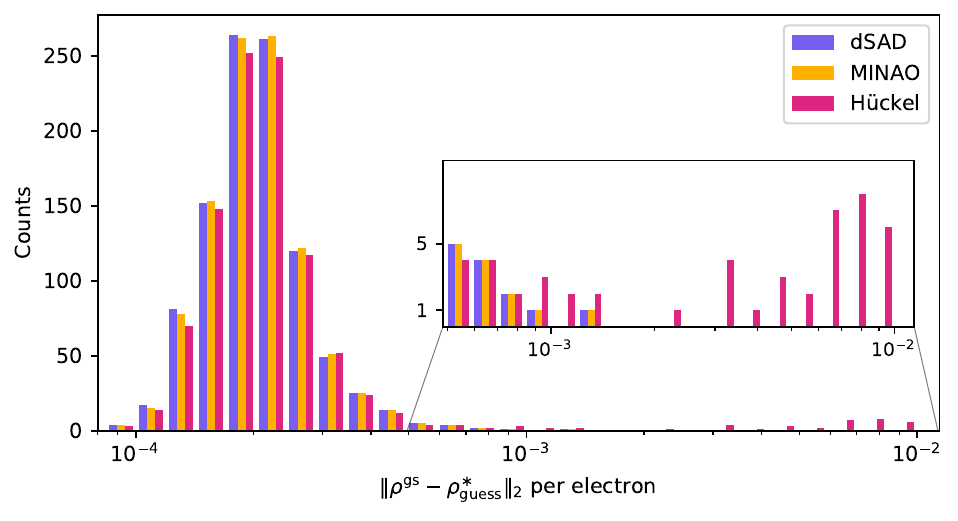}
    \caption[Distribution of STRUCTURES25 density optimization errors starting from different initializations.]{\textbf{Distribution of STRUCTURES25 density optimization errors starting from different initializations.} Shown is a histogram of $L_2$ density errors per electron for 1000 QM9 molecules for a dSAD, MINAO and Hückel initial guesses. The $\mathcal{O}(N)$ dSAD and $\mathcal{O}(N^3)$ MINAO perform similarly. Optimizations from the $\mathcal{O}(N^3)$ Hückel guess give rise to a small number of outliers which are highlighted in the inset.}
    \label{fig:ground_state_to_inits}
\end{figure}

On the same set of molecules, we compare density optimization results starting from the different guesses in Table~\ref{tab:denop_results_vary_inits} and Fig.~\ref{fig:ground_state_to_inits}. Using the identical hyperparameter configuration, both dSAD and MINAO guesses lead to all molecules converging, i.e.~achieving gradient norms below $10^{-4}~{\mathrm{Ha}}$, within 5000 optimization steps. Hückel initialization is worse, with only 79\% of molecules converging with default settings. Increasing the maximum number of iterations to 20k increases the convergence ratio to 85\%. Tuning of the momentum to a value of 0.77 pushes the Hückel convergence ratio to 96.7\% within 20k iterations.
The ability to converge to good solutions for all molecules from both dSAD and MINAO guesses using the same model is a testament to the generality of the STRUCTURES25 functional, as these two initializations differ considerably (cf.~Table~\ref{tab:initial_guess_errors}). 

Starting from different guesses, density optimizations with STRUCTURES25 converges to very similar solutions. Specifically, the predictions on 953 of the 1000 tested molecules differ at least an order of magnitude less to predictions from other initializations than they differ to the ground state label. The remainder exhibit density differences that are of the same order as the distance from the ground truth density. 

\begin{table}[ht]
    \centering
    \caption[STRUCTURES25 density optimization results with different initializations. ]{\textbf{STRUCTURES25 density optimization results with different initializations.} Compared are predicted ground state densities from dSAD, MINAO and Hückel guesses on 1000 molecules of the QM9 test set. For the  $L_2$ density errors per electron, we report the mean. For the number of required iterations to reach convergence we show the median [minimum, maximum] number of iterations. A hyphen ``--'' indicates lack of convergence within the allowed number of iterations. The first three rows correspond to density optimization configured by the default hyperparameter settings (see section~\ref{sec:density_optimization}). The last row (Hückel*) shows the results of an additional Hückel run with reduced momentum and greater number of allowed iterations.}
    \label{tab:denop_results_vary_inits}
    \vspace{10pt}
    \begin{tabular}{lccc} \thickhline
    Initial guess %
    & $\norm{\rho_{\mathrm{guess}} - \rho^{\mathrm{gs}}}_2 / \Ne \ (10^{-4})$ & Iterations & Converged \\ 
    &   &  median [min, max] & (\%) \\ \hline
    dSAD       %
    & \phantom{4}2.2 & \phantom{1}345 [226, 540]  & 100\phantom{.0}  \\
    MINAO     %
    & \phantom{4}2.2 & \phantom{1}500 [326, 635]  & 100\phantom{.0}  \\
    Hückel  %
    & 43.4 & \phantom{1}547 [323,\phantom{1.}--\phantom{1}]   & 79.0 \\
    Hückel* %
    & \phantom{4}5.7 & 1303 [839,\phantom{1.}--\phantom{1}]   & 96.7 \\ \thickhline
    \end{tabular}
\end{table}

\subsection{Ablation experiments}\label{sec:ablations}

The design of both the neural network architecture and the training procedure involves numerous choices.  This section details our ablation studies, performed to systematically evaluate the impact of key parameters and justify our final model configuration.  To minimize the influence of stochasticity, we report the best performance of three independent training runs (seeds) for each configuration, evaluating models based on the energy error unless otherwise specified.

\paragraph{Impact of perturbed training data:}

\begin{figure}
    \centering
    \includegraphics[width=0.7\linewidth]{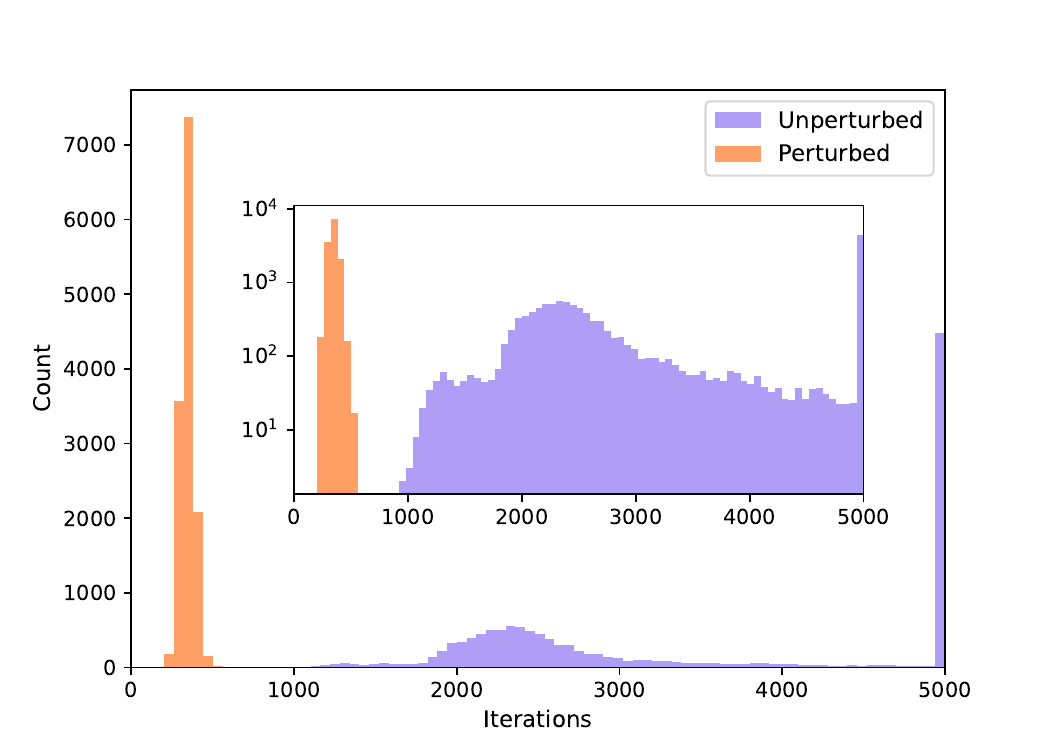}
    \caption{
    \textbf{Perturbation of the external potential for training data improves convergence in density optimization.} In density optimization, the model trained on conventional data generated without perturbing the external potential (purple) often does not converge while the model trained with data generated with our perturbation scheme consistently converges within a few hundred iterations (orange).
    }
    \label{fig:perturbation_convergence}
\end{figure}

Our primary contribution lies in generating training data that is both more diverse and more evenly distributed across the energy landscape as shown in Fig.~3C.  To quantify the benefit of this approach, we trained identical models on the QM9 dataset, once using only the standard Kohn-Sham SCF iterations (unperturbed) and once using our perturbed Fock matrix approach. Since the perturbed data has about 1.8 times the number of training labels, we increase the number of epochs for the non-perturbed model trainings to 161. Table~\ref{tab:perturbed_data_ablation} and Fig.~\ref{fig:perturbation_convergence} clearly demonstrate the advantages of perturbed data: the resulting model exhibits significantly improved convergence and achieves lower density errors. 

\begin{table}
    \centering
    \caption{
    \textbf{Perturbed training data ablation results on QM9.}}
    \vspace{10pt}
    \begin{tabular}{ccccccc}
        \thickhline
        Perturbed Data & $|\Delta E|$ & $|\Delta E|/ N_A$ & $\| \Delta \rho \|_2$ & $\| \Delta \rho \|_{2} / N_e$ & Convergence failures \\
        & (mHa) & (mHa) & & ($10^{-4}$) & (\%) \\
        \hline
        \checkmark  & 0.64 & 0.038 & 0.014 & \phantom{1} 2.1 & \phantom{0}0 \\
        $\times$    & 4.12 & 0.251 & 0.110 & 16.5 & 28 \\
        \thickhline
    \end{tabular}
    \vspace{10pt}
    \label{tab:perturbed_data_ablation}
\end{table}

\paragraph{Choice of energy target:}

Several options exist for the training target, each with distinct characteristics. One approach, termed ``delta learning,'' involves training on the difference between the non-interacting kinetic energy ($T_S$) and the APBEK approximation ($T_S - \mathrm{APBEK}$). This benefits from a smaller dynamic range in both energy and gradient values, which can simplify the learning process.  Another possible target is $E_\mathrm{TXC}$, which combines the non-interacting kinetic energy and the exchange-correlation energy.  A key advantage of this target is that it eliminates the need for numerical integration on a grid, significantly improving computational efficiency, particularly for larger molecules. Finally, we considered the total energy ($E_\mathrm{tot}$) as a target.  Ideally, this would allow the model to fully capture the energy minimum and its surrounding landscape, benefiting from the small gradient norms near the ground state.  However, accurately representing these small gradients proved challenging in practice.

Table~\ref{tab:energy_target_ablation} summarizes the results of training with each target.  The $T_S - \mathrm{APBEK}$ target, while viable, exhibits higher energy and density errors compared to the $E_\mathrm{TXC}$ target.  Furthermore, $T_S - \mathrm{APBEK}$ requires numerical integration on a grid for the evaluation of the APBEK functional and an XC functional, increasing the computational cost.  Models trained on $E_\mathrm{tot}$ fail to converge to meaningful densities, highlighting the difficulty of directly learning the total energy.  The superior performance and grid-free nature of $E_\mathrm{TXC}$ made it our target of choice.

\begin{table}
    \centering
    \caption{\textbf{Ablation of energy targets on QM9.}}
    \vspace{10pt}
    \begin{tabular}{cccccc} \thickhline
        Energy target & $|\Delta E|$ & $|\Delta E|/ N_A$ & $\| \Delta \rho \|_2$ & $\| \Delta \rho \|_{2} / N_e$ \\
        &  (mHa) &(mHa) & & ($10^{-4}$)\\
        \hline
         $E_\mathrm{TXC}$ & \phantom{118} 0.64 & \phantom{6} 0.038 & 0.014 & \phantom{27} 2.1 \\
         $T_\mathrm{S} - \mathrm{APBEK}$    & \phantom{118} 2.94 & \phantom{6} 0.178 & 0.037 & \phantom{27}5.7 \\
         $E_\mathrm{tot}$                   & 1183.88 & 62.314 & 1.858 & 279.2 \\
        \thickhline
    \end{tabular}
    \vspace{10pt}
    \label{tab:energy_target_ablation}
\end{table}

\paragraph{Tensorial vs.~scalar messages:}
We explore the impact of ``tensorial'' messages~\cite{lippmann2025beyond} in equivariant message passing based on local canonicalization, which allow the communication of non-scalar geometric information between nodes. We evaluate the performance of the standard Graphormer, which uses only scalar messages, against our modified version incorporating tensorial messages, as described in section~\ref{subsec:model_overview}.

Table~\ref{tab:tensorial_messages} shows results for the QM9 dataset. The additional geometric information improves the model, with the energy error improving slightly and the density error significantly. Networks trained with tensorial messages also showed lower gradient loss during training.

\paragraph{Number of Graphormer layers:}

The depth of the network, represented by the number of Graphormer layers, influences both the model's capacity and its computational cost.  We investigated the effect of varying the number of layers, with results presented in Table~\ref{tab:layer_ablation}.  Performance initially improves as the number of layers increases, allowing the model to capture more complex relationships. However, when no cutoff is used, we observe a significant degradation in performance beyond 4 layers, likely due to higher instability of the gradient produced by the network. This led us to select 4 layers as the optimal balance between expressivity and training stability for QM9, and 8 layers for QMugs.

\begin{table}[ht]
    \centering
    \caption{\textbf{Ablation of the number of Graphormer layers in the neural network on QM9.} For these experiments only a single seed was used.}
    \vspace{10pt}
    \begin{tabular}{cccccc}
        \thickhline
        \#Layers & \#Parameters & $|\Delta E|$ & $|\Delta E|/ N_A$ & $\| \Delta \rho \|_2$ & $\| \Delta \rho \|_{2} / N_e$ \\
        & ($10^{6}$) & (mHa) & (mHa) &  & ($10^{-4}$)\\
        \hline
         1 & \phantom{1} 8.1 & \phantom{00} 1.22 & \phantom{0} 0.074 & 0.025 & \phantom{0} 3.8 \\
         2 & 11.6 & \phantom{00} 0.75 & \phantom{0} 0.044 & 0.017 & \phantom{0} 2.6 \\
         3 & 15.1 & \phantom{00} 0.92 & \phantom{0} 0.053 & 0.015 & \phantom{0} 2.3 \\
         4 & 18.7 & \phantom{00} 0.64 & \phantom{0} 0.038 & 0.014 & \phantom{0} 2.1 \\
         6 & 25.8 & 439.29 & 19.871 & 0.198 & 29.4 \\
         8 & 32.9 & 322.92 & 14.214 & 0.097 & 14.4 \\
        
        \thickhline
    \end{tabular}
    \vspace{10pt}
    \label{tab:layer_ablation}
\end{table}

\paragraph{Fully connected vs. radial cutoff:}
The Graphormer architecture~\cite{ying2021transformers} was originally designed for fully connected graphs.  However, for scalability to larger systems, incorporating a radial cutoff is essential. In Table~\ref{tab:cutoff_ablation} we show that introducing a cutoff not only lowers computational cost but also yields smaller prediction errors.

\begin{table}[ht]
    \centering
    \caption{\textbf{Comparison of using a local vs.~fully-connected graph on the QMugs dataset.} The experiment was done after training on the mixed dataset of QM9 and QMugs molecules, without further fine-tuning.}
        \vspace{10pt}
    \begin{tabular}{cccccc}
    \thickhline
    local & $|\Delta E|$& $|\Delta E|/ N_A$ & $\| \Delta \rho \|_2$ & $\| \Delta \rho \|_{2} / N_e$ \\
    &  (mHa) &(mHa) && ($10^{-4}$)\\
    \hline
    \checkmark  & \phantom{5} 26 & 0.25 & 0.071 & 1.7 \\
    $\times$ & 580 & 8.80 & 0.195 & 7.3 \\
    \thickhline
    \end{tabular}
    \vspace{10pt}
    \label{tab:cutoff_ablation}
\end{table}

\subsection{Runtime comparison of KS-DFT and OF-DFT computations}\label{sec:runtime_comparison}
\begin{figure}[h]
    \centering
    \includegraphics[width=0.7\linewidth]{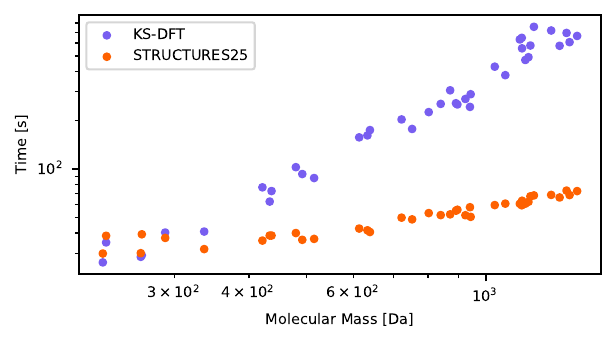}
    \caption{\textbf{Comparison of runtimes of KS-DFT and OF-DFT computations.}}
    \label{fig:timing}
\end{figure}

Figure \ref{fig:timing}
compares the runtimes of OF-DFT as embodied by STRUCTURES25 and KS-DFT as implemented in  \textsc{GPU4PySCF} \cite{li2024introducting,wu2024enhancing}, which can exploit the parallelism of the identical hardware. All computations were run on one Nvidia A100 GPU and 8 cores of an AMD EPYC 7452 processor. 
For the largest molecule of our QMUGS test set, STRUCTURES25 reaches a speedup of almost an order of magnitude. Importantly for future calculations on larger (e.g., biomolecular) systems, it shows favorable scaling with the size of the system.

\end{document}